\pdfoutput=1
\documentclass[utf8]{FrontiersinHarvard} % for articles in journals using the Harvard Referencing Style (Author-Date), for Frontiers Reference Styles by Journal: https://zendesk.frontiersin.org/hc/en-us/articles/360017860337-Frontiers-Reference-Styles-by-Journal
\usepackage{url,hyperref,lineno,microtype,subcaption}
\usepackage[onehalfspacing]{setspace}
\usepackage{graphicx}
\usepackage{graphics} % for pdf, bitmapped graphics files
\usepackage{amsmath} % assumes amsmath package installed
\usepackage{amssymb}  % assumes amsmath package installed
\usepackage{caption}
\usepackage{subcaption}
\def\mf{\mathbf}

\def\mc{\mathcal}

\def\zd{\mf z_D}
\def\za{\mf z_A}
\def\zb{\mf z_B}

\def\xd{\mf x_D}
\def\xa{\mf x_A}

\def\ta{\tau_A(\za,\Gamma)}
\def\td{\tau_D(\zd,\Omega)}
\def\taa{\tau_A(\za,\zb)}
\def\tdd{\tau_D(\zd,\zb)}

\def\bql{\begin{equation}}
\def\eql{\end{equation}}

\def\pp{p(\zd,\za,\zb)}

\def\pdstar{p(\zd,\za,\Omega^*,\Gamma)}
\def\pastar{p(\zd,\za,\Omega,\Gamma^*)}
\def\pdastar{p(\zd,\za,\Omega^*,\Gamma^*)}

\def\keyFont{\fontsize{8}{11}\helveticabold }
\def\firstAuthorLast{Lee {et~al.}} %use et al only if is more than 1 author
\def\Authors{Elijah S. Lee\,$^{1,*}$, Lifeng Zhou\,$^{2}$, Alejandro Ribeiro\,$^{1}$ and Vijay Kumar\,$^{1}$}
\def\Address{$^{1}$GRASP Laboratory, University of Pennsylvania, Philadelphia, PA, USA \\
$^{2}$ Department of Electrical and Computer Engineering, Drexel University, Philadelphia, PA, USA
}
\def\corrAuthor{Elijah S. Lee}

\def\corrEmail{elslee@seas.upenn.edu}

\begin{document}
\onecolumn
\firstpage{1}

\title[Graph Neural Networks for Decentralized Multi-Agent Perimeter Defense]{Graph Neural Networks for Decentralized Multi-Agent Perimeter Defense} 

\author[\firstAuthorLast ]{\Authors} %This field will be automatically populated
\address{} %This field will be automatically populated
\correspondance{} %This field will be automatically populated

\extraAuth{}% If there are more than 1 corresponding author, comment this line and uncomment the next one.
%\extraAuth{corresponding Author2 \\ Laboratory X2, Institute X2, Department X2, Organization X2, Street X2, City X2 , State XX2 (only USA, Canada and Australia), Zip Code2, X2 Country X2, email2@uni2.edu}

\maketitle

\begin{abstract}
%%% Leave the Abstract empty if your article does not require one, please see the Summary Table for full details.
% \section{}
In this work, we study the problem of decentralized multi-agent perimeter defense that asks for computing actions for defenders with local perceptions and communications to maximize the capture of intruders. One major challenge for practical implementations is to make perimeter defense strategies scalable for large-scale problem instances. To this end, we leverage graph neural networks (GNNs) to develop an imitation learning framework that learns a mapping from defenders' local perceptions and their communication graph to their actions. The proposed GNN-based learning network is trained by imitating a centralized expert algorithm such that the learned actions are close to that generated by the expert algorithm. We demonstrate that our proposed network performs closer to the expert algorithm and is superior to other baseline algorithms by capturing more intruders. Our GNN-based network is trained at a small scale and can be generalized to large-scale cases. We run perimeter defense games in scenarios with different team sizes and configurations to demonstrate the performance of the learned network.
\tiny
 \keyFont{ \section{Keywords:} graph neural networks, perimeter defense, multi-agent systems, perception-action-communication loops, imitation learning} %All article types: you may provide up to 8 keywords; at least 5 are mandatory.
\end{abstract}

% SECTIONS
%%%%%%%%%%%%%%%%%%%%%%%%%%%%%%%%%%%%%%%%%%%%%%%%%%%%%%%%%%%%%%%%%%%%%%%%%%%%%%%%
\section{Introduction}
\label{sec:intro}

The problem of perimeter defense games considers a scenario where the defenders are constrained to move along a perimeter and try to capture the intruders while the intruders aim to reach the perimeter without being captured by the defenders~\citep{shishika2020review}. A number of previous works have solved this problem with engagements on a planar game space~\citep{shishika2018local, chen2021optimal}. However, in the real world, the perimeter may be represented by a three-dimensional shape as the players (e.g., defenders and intruders) may have the ability to perform three-dimensional motions. For example, a perimeter of a building that defenders aim to protect can be enclosed by a hemisphere. As a result, the defender robots should be able to move in three-dimensional space. For example, aerial robots have been well studied in various settings~\citep{chen2020sloam, nguyen2019mavnet, lee2016drone, lee2020experimental}, and all these settings can be real-world use-cases for perimeter defense. For instance, intruders try to attack a military base in the forest and defenders aim to capture the intruders.

In this work, we tackle the perimeter defense problem in a domain where multiple agents collaborate to accomplish a task. Multi-agent collaboration has been explored in many areas including environmental mapping~\citep{liu2022active, thrun2000real}, search and rescue~\citep{miller2020mine, baxter2007multi}, target tracking~\citep{ge2022vision, lee2022learning}, on-demand wireless infrastructure~\citep{mox2020mobile}, transportation~\citep{xu2022modular, ng2022takes}, and multi-agent learning~\citep{kim2021policy}. Our approach employs a team of robots that work collectively towards a common goal of defending a perimeter. We focus on developing decentralized strategies for a team of defenders for various reasons: (i) the teammates can be dynamically added or removed without disrupting explicit hierarchy; (ii) the centralized system may fail to cope with the high dimensionality of a team's joint state space; and (iii) the defenders have a limited communication range and can only communicate locally. 

To this end, we aim to develop a framework where a team of defenders collaborates to defend the perimeter using decentralized strategies based on local perceptions and communications. Specifically, we explore learning-based approaches to learn policies by imitating expert algorithms such as the maximum matching algorithm~\citep{chen2014multiplayer}. Maximum matching algorithm that runs the exhaustive search to find the best policy is very computationally intensive at large scales since this approach is combinatorial in nature and assumes global information. We utilize GNN as the learning paradigm and demonstrate that the trained network can perform close to the expert algorithm. GNNs have decentralized communication architecture that capture the neighboring interactions and transferability that allows for generalization to previously unseen scenarios~\citep{ruiz2021graph}. We demonstrate that our proposed GNN-based network can be generalized to large scales in solving multi-robot perimeter defense games.

With this insight, we make the following primary contributions in this paper:

\textbf{Framework for decentralized perimeter defense using graph neural networks.} We propose a novel learning framework that utilizes a graph-based representation for the perimeter defense game. To the best of our knowledge, we are the first to solve the decentralized hemisphere perimeter defense problem by learning decentralized strategies via graph neural networks.
    
\textbf{Robust perimeter defense performance with scalability.} We demonstrate that our methods perform close to an expert policy (i.e., maximum matching algorithm~\cite{chen2014multiplayer}) and are superior to other baseline algorithms. Our proposed networks are trained at a small scale and can be generalized to large scales.

%%%%%%%%%%%%%%%%%%%%%%%%%%%%%%%%%%%%%%%%%%%%%%%%%%%%%%%%%%%%%%%%%%%%%%%%%%%%%%%%
\section{Related Work}
\label{sec:related}

\subsection{Perimeter Defense} 
In a perimeter defense game, defenders aim to capture intruders by moving along a perimeter while intruders try to reach the perimeter without being captured by the defenders. We refer to~\citep{shishika2020review} for a detailed survey. Many previous works dealt with engagements on a planar game space~\citep{shishika2018local, Macharet2020Adaptive, chen2021optimal, bajaj2021competitive, hsu2022model}. For example, a cooperative multiplayer perimeter-defense game was solved on a planar game space in~\citep{shishika2018local}. In addition, an adaptive partitioning strategy based on intruder arrival estimation was proposed in~\citep{Macharet2020Adaptive}.
Later, a formulation of the perimeter defense problem as an instance of the flow networks was proposed in~\citep{chen2021optimal}. Further, an engagement on a conical environment was discussed in~\citep{bajaj2021competitive}, and a model with heterogeneous teams was addressed in~\citep{hsu2022model}. 

High-dimensional extensions of the perimeter defense problem have been recently explored in~\citep{lee2021guarding, yan2022matching, lee2020perimeter, lee2021defending, lee2022vision}. For example, \citet{lee2021guarding} analyzed the two-player differential game of guarding a closed convex target set from an attacker in high-dimensional Euclidean spaces. \citet{yan2022matching} studied a 3D multiplayer reach-avoid game where multiple pursuers defend a goal region against multiple evaders. \citet{lee2020perimeter, lee2021defending, lee2022vision} considered a game played between aerial defender and ground intruder.

All of the aforementioned works focus on solving centralized perimeter defense problems, which assume that players have global knowledge of other players' states. However, decentralized control becomes a necessity as we reach a large number of players. To remedy this problem, \citet{velhal2022decentralized} formulated the perimeter defense game into a decentralized multi-robot spatio-temporal multitask assignment problem on the perimeter of a convex shape. \citet{paulos2019decentralization} proposed neural network architecture for training decentralized agent policies on the perimeter of a unit circle, where defenders have simple binary action spaces. Different from the aforementioned works, we focus on the high-dimensional perimeter, specialized to a hemisphere, with continuous action space. We solve multi-agent perimeter defense problems by learning decentralized strategies with graph neural networks.

\subsection{Graph Neural Networks}
We leverage graph neural networks as the learning paradigm because of their desirable properties of decentralized architecture that captures the interactions between neighboring agents and transferability that allows for generalization to previously unseen cases~\citep{Gama19-Architectures, ruiz2021graph}. In addition, GNNs have shown great success in various multi-robot problems such as formation control~\citep{Tolstaya19-Flocking}, path planning~\citep{li2020message}, task allocation~\citep{wang2020learning}, and multi-target tracking~\citep{zhou2021graph,sharma2022d2coplan}. Particularly,~\citet{Tolstaya19-Flocking} utilized a GNN to learn a decentralized flocking behavior for a swarm of mobile robots by imitating a centralized flocking controller with global information. Later,~\citet{li2020message} implemented GNNs to find collision-free paths for multiple robots from start positions to goal positions in obstacle-rich environments. They demonstrated that their decentralized path planner achieves a near-expert performance with local observations and neighboring communication only, which can also be generalized to larger networks of robots. The GNN-based approach was also employed to learn solutions to the combinatorial optimization problems in a multi-robot task scheduling scenario~\citep{wang2020learning} and multi-target tracking scenario~\citep{zhou2021graph,sharma2022d2coplan}.

%%%%%%%%%%%%%%%%%%%%%%%%%%%%%%%%%%%%%%%%%%%%%%%%%%%%%%%%%%%%%%%%%%%%%%%%%%%%%%%%
\section{Problem Formulation}
\label{sec:problem}

\subsection{Motivation}\label{sec:motivation}
Perimeter defense is a relatively new field of research that has been explored recently. One particular challenge is that the high-dimensional perimeters add spatial and algorithmic complexities for defenders to execute their optimal strategies. Although many previous works considered engagements on a planar game space and derived optimal strategies in 2D motions, the extension towards high-dimensional spaces is unavoidable for practical applications of perimeter defense games in real-world scenarios. For instance, a perimeter of a building that defenders aim to protect can be enclosed by a generic shape, such as a hemisphere. Since defenders cannot pass through the building and are assumed to be close to the building at any time, they are employed to move along the surface of the dome, which leads to the ``hemisphere perimeter defense game.'' The intruder is moving on the base plane of the hemisphere, which implies a constant altitude during moving. The movement of the intruder is constrained to 2D since it is assumed that intruders may want to stay low in altitude to hide from the defenders in the real world.

It is worth noting that the hemisphere defense problem is more challenging to solve than a problem where both agents are allowed to freely move in a 3D space. There were previous works in which both defenders and intruders could move in 3-dimensional spaces~\citep{yan2022matching,yan2019construction,yan2020guarding}. In all cases, the authors were able to explicitly derive the optimal solutions even in multi-robot scenarios. Although our problem limits the dynamics of the defenders to the surface of the hemisphere, these constraints make the finding of an optimal solution intractable and challenging.

\subsection{Hemisphere Perimeter Defense}
\begin{figure}[b!]
\begin{center}
\includegraphics[height=7cm]{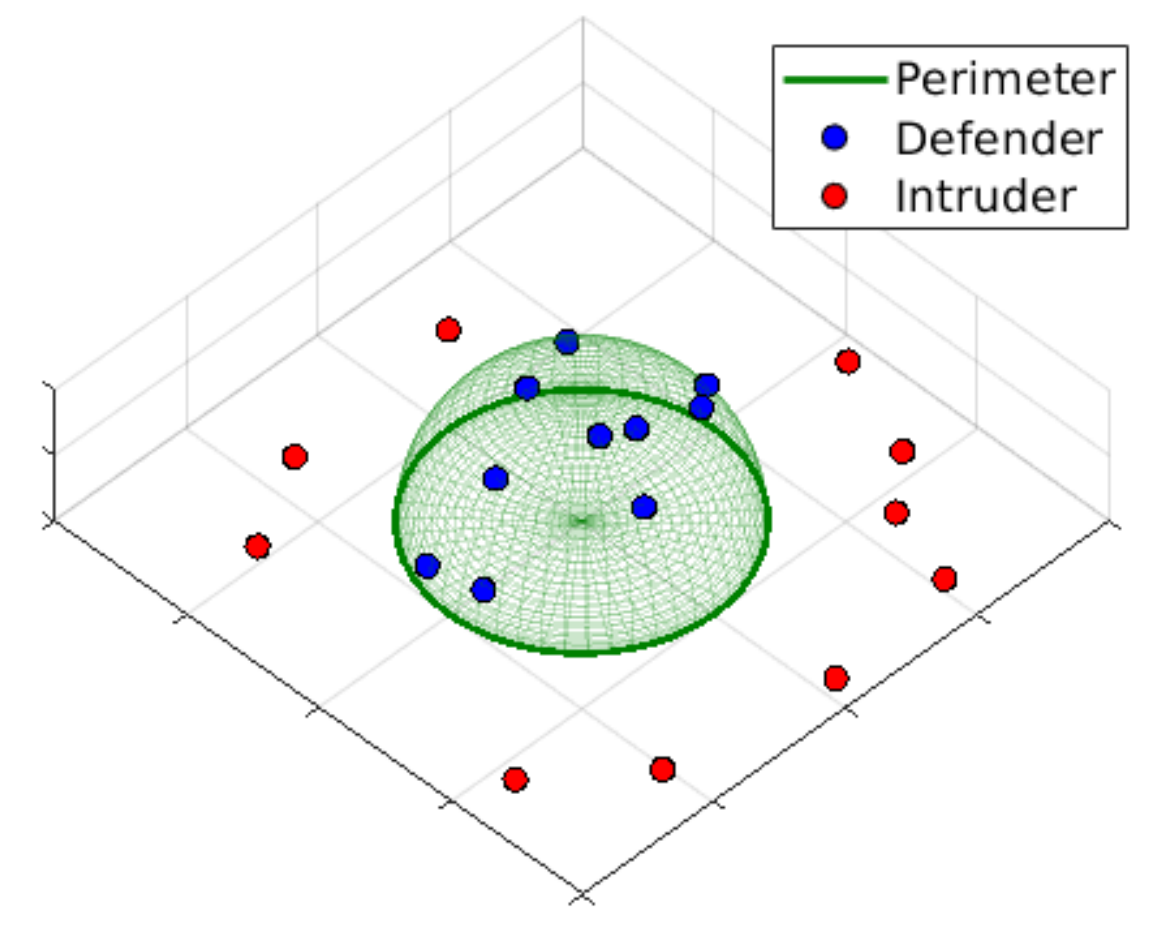}
\end{center}
\caption{Instance of 10 vs. 10 perimeter defense. Defenders are constrained to move on the surface of the dome while intruders are constrained to move on the ground plan.}
\label{fig:problem}
\end{figure}

We consider a hemispherical dome with radius of $R$ as perimeter. The hemisphere constraint is for the defender to safely move around the perimeter (e.g. building). In this game, consider two sets of players: $\textbf{D}=\{D_i\}^N_{i=1}$ denoting $N$ defenders, and $\textbf{A}=\{A_j\}^N_{j=1}$ denoting $N$ intruders. A defender $D_i$ is constrained to move on the surface of the dome while an intruder $A_j$ is constrained to move on the ground plane. We will drop the indices $i$ and $j$ when they are irrelevant. An instance of 10 vs. 10 perimeter defense is shown in Figure~\ref{fig:problem}. The positions of the players in spherical coordinates are: $\zd=[\psi_D,\phi_D,R]$ and $\za=[\psi_A,0,r]$, where $\psi$ and $\phi$ are the azimuth and elevation angles, which gives the relative position as: $\mf z \triangleq [\psi,\phi,r]$, where $\psi\triangleq \psi_A-\psi_D$ and $\phi\triangleq \phi_D$. The positions of the players can also be described in Cartesian coordinates as: $\xd$ and $\xa$. All agents move at unit speed, defenders capture intruders by closing within a small distance $\epsilon$, and both defender and intruder are consumed during capture. An intruder wins if it reaches the perimeter (i.e., $r(t_f)=R$) at time $t_f$ without being captured by any defenders (i.e., $||\mf x_{A_i}(t) - \mf x_{D_j}(t)||>\epsilon, \forall D_j \in \mathbf{D}, \forall t < t_f$). A defender wins by capturing an intruder or preventing it from scoring indefinitely (i.e., $\phi(t)=\psi(t)=0$, $r(t)>R$). The main interest of this work is to maximize the number of captures by defenders, given a set of initial configurations. 

\begin{figure}[t!]
\begin{center}
\includegraphics[height=7cm]{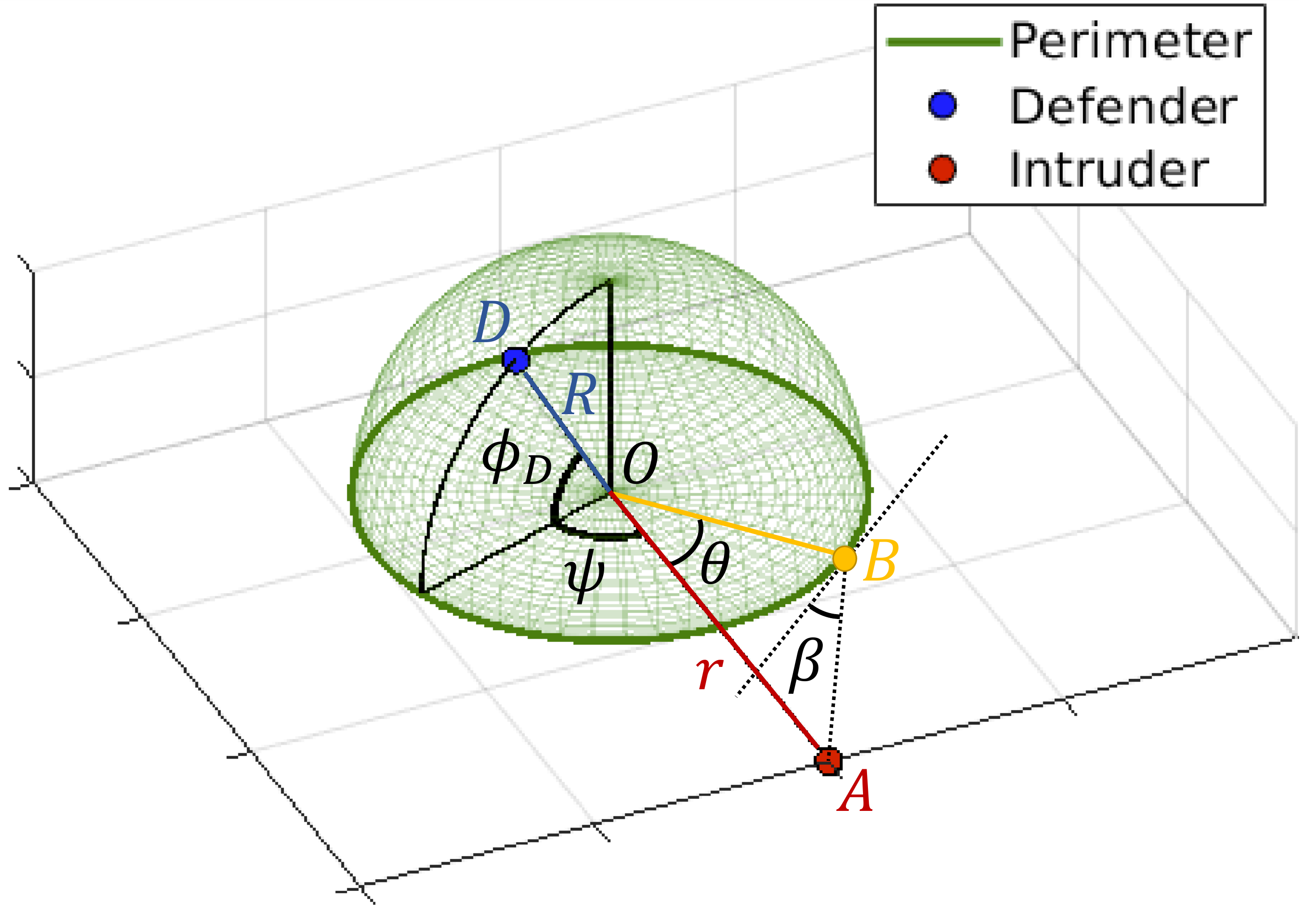}
\end{center}
\caption{Coordinates and relevant variables in the 1 vs. 1 hemisphere defense game.}
\label{fig:hemisphere}
\end{figure}

\subsection{Optimal Breaching Point}\label{sec:optimal}
Given $\zd$, $\za$, we call $\textit{breaching point}$ as a point on the perimeter at which the intruder tries to reach the target, as shown $B$ in Figure~\ref{fig:hemisphere}. We call the azimuth angle that forms the breaching point as \textit{breaching angle}, denoted by $\theta$, and call the angle between $(\mf z_A - \mf z_B)$ and the tangent line at $B$ as \textit{approach angle}, denoted by $\beta$. It is proved in~\citep{lee2020perimeter} that given the current positions of defender $\mf z_D$ and intruder $\mf z_A$ as point particles, there exists a unique breaching point such that the optimal strategy for both defender and intruder is to move towards it, known as \textit{optimal breaching point}. The breaching angle and approach angle corresponding to the optimal breaching point are known as \textit{optimal breaching angle}, denoted by $\theta^*$, and \textit{optimal approach angle}, denoted by $\beta^*$. As stated in \citep{lee2020perimeter}, although there exists no closed-form solution for $\theta^*$ and $\beta^*$, they can be computed at any time by solving two governing equations:

\begin{equation}
\beta^* =  \cos^{-1}\left(\nu\frac{\cos{\phi_D}\sin{\theta^*}}{\sqrt{1-\cos^2{\phi_D}\cos^2{\theta^*}}}\right)
\label{eq:beta}
\end{equation}
and
\begin{equation}
\theta^* = \psi-\beta^*+\cos^{-1}\left(\frac{\cos\beta^*}{r}\right) \label{eq:theta}
\end{equation}

\subsection{Target Time and Payoff Function}\label{sec:payoff}
We call the \textit{target time} as the time to reach $B$ and define $\tdd$ as the \textit{defender target time}, $\taa$ as the \textit{intruder target time}, and the following as \textit{payoff} function:

\bql
\pp = \tdd -\taa \label{eq:payoff}
\eql   

The defender reaches $B$ faster if $p<0$ and the intruder reaches $B$ faster if $p>0$. Thus, the defender aims to minimize $p$ while the intruder aims to maximize it.

\subsection{Optimal Strategies and Nash Equilibrium  \label{sec:nash}}
It is proven in \citep{lee2020perimeter} that the optimal strategies for both defender and intruder are to move towards the optimal breaching point at their maximum speed at any time. Let $\Omega$ and $\Gamma$ be the continuous $v_D$ and $v_A$ that lead to $B$ so that $\td \triangleq \tdd$ and $\ta \triangleq \taa$, and let $\Omega^*$ and $\Gamma^*$ be the optimal strategies that minimize $\td$ and $\ta$, respectively, then the optimality in the game is given as a Nash equilibrium:
\bql
\pdstar\leq\pdastar\leq\pastar \label{eq:nash}
\eql

\subsection{Problem Definition}

To maximize the number of captures during $N$ vs. $N$ defense, we first recall the dynamics of a 1 vs. 1 perimeter defense game. It is proven in~\citep{lee2020perimeter} that the best action for the defender in one-on-one game is to move towards the \textit{optimal breaching point} (defined in Section~\ref{sec:optimal}). The defender reaches the optimal breaching point faster than the intruder does if \textit{payoff} $p$ (defined in Section~\ref{sec:payoff}) is negative, and the intruder reaches faster if $p>0$. From this, we infer that maximizing the number of captures in $N$ vs. $N$ defense is the same as finding a matching between the defenders and intruders so that the number of the negative payoff of assigned pairs is maximized. In an optimal matching, the number of negative payoffs stays the same throughout the overall game since the optimality in each game of defender-intruder pairs is given as a \textit{Nash equilibrium} (see Section~\ref{sec:nash}).

The expert assignment policy is a \textit{maximum matching}~\citep{shishika2018local, chen2014multiplayer}. To execute this algorithm, we generate a bipartite graph with $\textbf{D}$ and $\textbf{A}$ as two sets of nodes (i.e., $\mc{V}=\{1,2,..,N\}$), and define the potential assignments between defenders and intruders as the edges. For each defender/node $D_i$ in $\textbf{D}$, we find all the intruders/nodes $A_j$ in $\textbf{A}$ that are sensible by the defender and compute the corresponding payoffs $p_{ij}$ for all the pairs. We say that $D_i$ is \textit{strongly assigned} to $A_j$ if $p_{ij}<0$. Using the edge set $\mathcal{E}$ given by maximum matching, we can maximize the number of strongly assigned pairs. For uniqueness, we choose a matching that minimizes the \textit{value of the game}, which is defined as
\begin{align}\label{eqn:V}
    V = \sum_{(D_i,A_j)\in \mathcal{E}^*} p_{ij},
\end{align}
where $\mathcal{E}^*$ is the subset of $\mathcal{E}$ with negative payoff (i.e. $\mathcal{E}^*= \{(D_i,A_j)\in\mathcal{E} \mid p_{ij}<0\}$). This unique assignment ensures that the number of captures is maximized at the earliest possible. However, running the exhaustive search using maximum matching algorithm can be very expensive as the team size increases. This method is combinatorial in nature and assumes centralized information with full observability. Instead, we aim to find decentralized strategies that uses local perceptions $\{\mc{Z}_i\}_{i\in\mc{V}}$ (see Section~\ref{subsec:perception}). To this end, we formalize the main problem of this paper as follows.

\textsc{Problem 1} (Decentralized Perimeter Defense with Graph Neural Networks). \textit{
Design a GNN-based learning framework to learn a mapping $\mc{M}$ from the defenders' local perceptions $\{\mc{Z}_i\}_{i\in\mc{V}}$ and their communication graph $\mc{G}$ to their actions $\mc{U}$, i.e., $\mc{U} = \mc{M}(\{\mc{Z}_i\}_{i\in\mc{V}}, \mc{G})$, such that $\mc{U}$ is as close as possible to action set $\mc{U}^\texttt{g}$ selected by a centralized expert algorithm.}
\label{prob:learning}

We describe in detail our learning architecture for solving Problem 1 in the following section.

%%%%%%%%%%%%%%%%%%%%%%%%%%%%%%%%%%%%%%%%%%%%%%%%%%%%%%%%%%%%%%%%%%%%%%%%%%%%%%%%
\section{Method}
\label{sec:method}

\begin{figure}[t!]
    % \centering
    \includegraphics[width=1\textwidth]{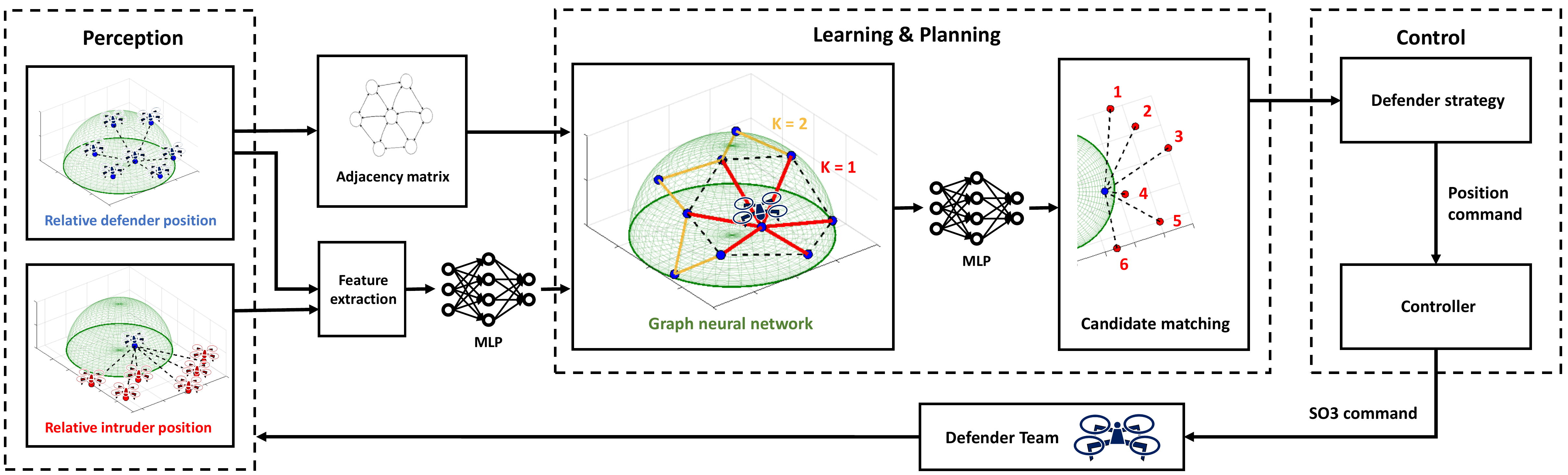}
    \caption{Overall framework. Perception module collects local information. Learning \& Planning module processes the collected information using GNN through $K$-hop neighboring communications. Control module computes the optimal strategies and executes the controller to close the loop.}
    \label{fig:framework}
\end{figure}

In this paper, we learn decentralized strategies for perimeter defense using graph neural networks. Inference of our approach is in real-time, which is scalable to a large number of agents. We use an expert assignment policy to train a team of defenders who share information through communication channels. In Section~\ref{subsec:perception}, we introduce the perception module for processing the features that are input to GNN. Learning the decentralized algorithm through GNN and planning the candidate matching for the defenders are discussed in Section~\ref{subsec:learnandplan}. The control of the defender team is explained in Section~\ref{subsec:control}, and the training procedure is detailed in Section~\ref{subsec:training}. The overall framework is shown in Figure~\ref{fig:framework}. For the choice of architecture, we decouple the control module from the learning framework since directly learning the actions is unnecessary. Learning an assignment between agents is sufficient, and the best actions can be computed by the optimal strategies (Section~\ref{sec:nash}).

\subsection{Perception}\label{subsec:perception}

In this section, we assume $N$ aerial defenders and $N$ ground intruders. Each defender $D_i$ is equipped with a sensor and faces outwards the perimeter with a field of view $FOV$. The defenders' horizontal field of view $FOV$ is chosen as $\pi$ assuming a fisheye-type camera. 

\subsubsection{Intruder features}
For each $i$, a defender observes the set of intruders $A_j$, and the relative positions in spherical coordinates between $D_i$ and $A_j$ are represented by $\mc{Z}_i^A = \{\mf z_{ij}^A\}_{j\in\mathit{N}_A^f}$ where $\mathit{N}_A^f$ is the number of intruder features. The number of input features $\mathit{N}_A^f$ and $\mathit{N}_D^f$ are selected as the fixed number of closest detected and neighboring agents, respectively. Although a defender can detect any number of intruders within the sensing range, a fixed number of detections is selected so that the system is scalable. In a decentralized setting, a defender should be able to decide its action based on its local perceptions. We experimentally chose the fixed number as 10 since an expert algorithm (i.e., the maximum matching) would always assign a defender to a robot among the 10 closest intruders. 

\subsubsection{Defender features}
To make the system scalable, we build communication with a fixed number of closest defenders. Each defender $D_i$ communicates with nearby defenders $D_j$ within its communication range $r_c$. For each $i$, the relative positions between $D_i$ and $D_j$ are represented by $\mc{Z}_i^D = \{\mf z_{ij}^D\}_{j\in\mathit{N}_D^f}$ where $\mathit{N}_D^f$ is the number of defender features. The selected number was 3 since communicating with many other robots would allow every defender to have full information of the environment (i.e., centralized) and 3 is the minimum number that the robots can collect information in every direction if we assume robots are scattered. If there are fewer than 10 detected intruders or 3 neighboring defenders, we hand over dummy values to fill up the perception input matrix. It is important to keep the input features constant since neural networks cannot handle varying feature sizes.

\subsubsection{Feature extraction}
Feature extraction is performed by concatenating the relative positions of observed intruders and communicated defenders, forming the local perceptions $\mc{Z}_i = \{\mc{Z}_i^A, \mc{Z}_i^D\}$. The extracted features are fed into a multi-layer perceptron (MLP) to generate the post-processed feature vector $\mf x_i$, which will be exchanged among neighbors through communications.

\subsection{Learning \& Planning}\label{subsec:learnandplan}
We employ graph neural networks with $K$-hop communications. Defenders communicate their perceived features with neighboring robots. The communication graph $\mc G$ is formed by connecting the nearby defenders within the communication range $r_c$, and the resulted adjacency matrix $\mf S$ is given to the graph neural networks.

\subsubsection{Graph Shift Operation} We consider each defender $i, i\in \mathcal{V}$ has a feature vector $\mathbf{x}_i \in \mathbb{R}^F$, indicating the post-processed information from $D_i$. By collecting the feature vectors $\mathbf{x}_i$ from all defenders, we have the feature matrix for the defender team $\mf D$ as: 
\begin{equation} \label{eqn:featureMatrix}
    \mathbf{X} 
    = \begin{bmatrix}
        \mathbf{x}_1^{\mathsf{T}} \\
        \vdots \\
        \mathbf{x}_N^{\mathsf{T}}
      \end{bmatrix} = [\mathbf{x}^1, \cdots, \mathbf{x}^F] \in \mathbb{R}^{N\times F},
\end{equation}
where $\mathbf{x}^f \in  \mathbb{R}^N, f \in [1, \cdots, F]$ is the collection of the feature $f$ across all defenders; i.e., $\mathbf{x}^f = [\mathbf{x}_1^f, \cdots, \mathbf{x}_N^f]^{\mathsf{T}}$ with $\mathbf{x}_i^f$ denoting the feature $f$ of $D_i, i\in\mathcal{V}$. We conduct \textit{graph shift operation} for each $D_i$ by a linear combination of its neighboring features, i.e., $\sum_{j\in \mathcal{N}_i} \mathbf{x}_j$. Hence, for all defenders $\mf D$ with graph $\mathcal{G}$, the feature matrix $\mathbf{X}$ after the shift operation becomes $\mathbf{S} \mathbf{X}$ with:   
\begin{equation} \label{eqn:graphShift}
    [\mathbf{S} \mathbf{X}]_{if} 
        = \sum_{j = 1}^{N} [\mathbf{S}]_{ij} [\mathbf{X}]_j^f
        = \sum_{j \in \mathcal{N}_{i}}
           s_{ij} \mathbf{x}_j^f, 
\end{equation}
Here, the adjacency matrix $\mathbf{S}$ is called the \emph{Graph Shift Operator} (GSO)~\citep{Gama19-Architectures}.

\subsubsection{Graph Convolution} With the shift operation, we define the \textit{graph convolution} by a linear combination of the \textit{shifted features} on graph $\mathcal{G}$ via $K$-hop communication exchanges \citep{Gama19-Architectures,li2019graph}: 
\begin{equation} \label{eqn:graphConvolution}
    \mathcal{H}(\mathbf{X}; \mathbf{S}) = \sum_{k=0}^{K} \mathbf{S}^{k} \mathbf{X} \mathbf{H}_{k},
\end{equation}
where $\mathbf{H}_{k} \in \mathbb{R}^{F \times G}$ represents the coefficients combining $F$ features of the defenders in the shifted feature matrix $\mathbf{S}^{k} \mathbf{X}$, with $F$ and $G$ denoting the input and output dimensions of the graph convolution. Note that, $\mathbf{S}^{k} \mathbf{X} = \mathbf{S}(\mathbf{S}^{k-1} \mathbf{X}) $ is computed by means of $k$ communication exchanges with $1$-hop neighbors.

\subsubsection{Graph Neural Network} Applying a point-wise non-linearity $\sigma: \mathbb{R} \to \mathbb{R}$ as the activation function to the graph convolution (Eq.~\ref{eqn:graphConvolution}), we define \textit{graph perception} as: 
\begin{equation} \label{eqn:graphPerception}
    \mathcal{H}(\mathbf{X}; \mathbf{S}) = \sigma(\sum_{k=0}^{K} \mathbf{S}^{k} \mathbf{X} \mathbf{H}_{k}).
\end{equation}

Then, we define a GNN module by cascading $L$ layers of graph perceptions (Eq.~\ref{eqn:graphPerception}):
\begin{equation} \label{eqn:convGNN}
    \mathbf{X}^{\ell} = \sigma \big[ \mathcal{H}^{\ell}(\mathbf{X}^{\ell-1};\mathbf{S}) \big] \quad \text{for} \quad \ell = 1,\cdots,L,
\end{equation}
where the output feature of the previous layer $\ell-1$, $\mathbf{X}^{\ell-1} \in \mathbb{R}^{N \times F^{\ell-1}}$, is taken as input to the current layer $\ell$ to generate the output feature of layer $l$, $\mathbf{X}^{\ell}$. Recall that the input to the first layer is $\mathbf{X}^{0} = \mathbf{X}$ (Eq.~\ref{eqn:featureMatrix}). The output feature of the last layer $\mathbf{X}^{L} \in \mathbb{R}^{N \times G}$, obtained via $K$-hop communications, represents the exchanged and fused information of the defender team $\mf D$.

\subsubsection{Candidate matching}
The output of the GNN, which represents the fused information from the $K$-hop communications, is then processed with another MLP to provide a candidate matching for each defender. Figure~\ref{fig:framework} shows a candidate matching instance if $\mathit{N}_A^f=6$. Given a defender $D_i$, we find the $\mathit{N}_A^f$ closest intruders and number them from 1 to $\mathit{N}_A^f$ clockwise. The main reason for numbering the nearby intruders clockwise is to interpret the feature outputs from our networks in identifying which intruders would be matched with which defenders. We could number them counterclockwise or in any arbitrary order. Since each defender learns decentralized strategies, it needs to specify an intruder to capture given its local perception. There are no global IDs for the intruders so without loss of generality we simply assign the IDs clockwise. The output from the multi-layer perceptron is an assignment likelihood $\mc L$, which presents the probabilities of $\mathit{N}_A^f$ intruder candidates' likelihood to be matched with the given defender. For instance, an expert assignment likelihood $L_i^g$ for $D_i$ in Figure~\ref{fig:framework} would be $[0.01,0.01,0.95,0.01,0.01,0.01]$ if the third intruder (i.e., $A_3$) is matched with $D_i$ by the expert policy (i.e., maximum matching). The planning module selects the intruder candidate $A_j$ so that the matching pair $(D_i, A_j)$ would resemble the expert policy with the highest probability. It is worth noting that our approach renders a decentralized assignment policy given that only neighboring information is exchanged.

\subsubsection{Permutation Equivalence}\label{sec:ordering}

\begin{figure}[t!]
    \centering
    \includegraphics[width=1\textwidth]{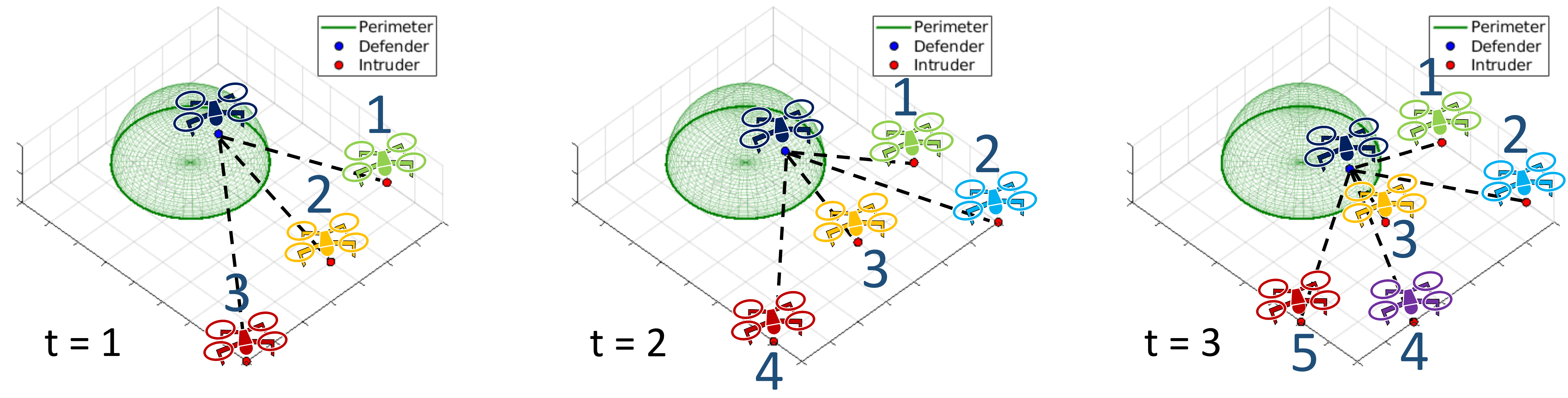}
    \caption{Instance of perimeter defense game at different time stamps. The plots focus on a single defender and its local perceptions.}
    \label{fig:ordering}
\end{figure}

It is worth noting that our proposed GNN-based learning approach is scalable due to permutation equivalence. This means that given a decentralized defender, it should be able to decide the action based on local perceptions that consist of an arbitrary number of unnumbered intruders. An instance of a perimeter defense game is illustrated to show this property in Figure~\ref{fig:ordering}. The plots focus on a single defender and intruders are gradually approaching the perimeter as time passes by. The same intruders are colored in the same color across different time stamps. Notice that a new light-blue intruder enters into the field of view of the defender at $t=2$, and a purple intruder begins to appear at $t=3$. Although an arbitrary number of intruders are detected at each time, our system gives IDs to intruders shown as blue numbers in Figure~\ref{fig:ordering}. We number them clockwise but could have done differently in any permutation (e.g., counterclockwise) because graph neural networks perform label-independent processing. The reason for the numbering is to specify which intruders would be matched with which defenders from the network outputs. Without loss of generality, we assign the IDs clockwise but we note that these IDs are arbitrary since the IDs can change at different stamps. For instance, the yellow intruder ID is 2 at $t=1$ but becomes 3 at $t=2,3$. Similarly, the red intruder ID is 3 at $t=1$ but changes to 4 at $t=2$ and 5 at $t=3$. In this way, we accommodate an arbitrary amount of intruders and thus our system is permutation equivalent.

\subsection{Control}\label{subsec:control}
The output from the Section~\ref{subsec:learnandplan} is inputted to the defender strategy module in Figure~\ref{fig:framework}. This module handles all the matched pairs $(D_i, A_j)$ and computes the optimal breaching points for each of the one-on-one hemisphere perimeter defense games (see Section~\ref{sec:optimal}). The defender strategy module collectively outputs the position commands, which are towards the direction of the optimal breaching points. The SO(3) command~\citep{mellinger2011minimum} that consists of thrust and moment to control the robot at a low level is then passed to the defender team $\textbf{D}$ for control. The state dynamics for the defender-intruder pair is detailed in~\citep{lee2020perimeter}. The defenders move based on the commands to close the perception-action loop. Notably, the expert assignment likelihood $\mc L^g$ would result in the expert action set $\mc U^g$ (defined in Problem 1).

\subsection{Training Procedure}\label{subsec:training}
To train our proposed networks, we use imitation learning to mimic an expert policy given by maximum matching (explained in Section~\ref{sec:problem}), which provides the optimal assignment likelihood $\mc L^g$ (described in Section~\ref{subsec:learnandplan}) given the defenders' local perceptions $\{\mc{Z}_i\}_{i\in\mc{V}}$ and the communication graph $\mc G$. The training set $\mc D$ is generated as a collection of these data: $\mc D = \{(\{\mc{Z}_i\}_{i\in\mc{V}}, \mc G, \mc L^g)\}$. We train the mapping $\mc M$ (defined in Problem 1) to minimize the cross-entropy loss between $\mc L^g$ and $\mc L$. We show that the trained $\mc M$ provides $\mc U$ that is close to $\mc U^g$. The number of learnable parameters in our networks is independent of the number of team sizes $N$. Therefore, we can train our networks on a small scale and generalize our model to large scales, given that defenders at any scale learn decentralized strategies based on the local perception of fixed numbers of agents.

\subsubsection{Model Architecture}\label{sec:modelArchitecture}
Our model architecture consists of a 2-layer MLP with 16 and 8 hidden layers to generate the post-processed feature vector $\mf x_i$, a 2-layer GNN with 32 and 128 hidden layers to exchange the collected information from defenders, and a single-layer MLP to produce an assignment likelihood $\mc L$. The layers in MLP and GNN are followed by ReLU.

\subsubsection{Graph Neural Networks Details}\label{sec:graphDetail}
In implementing graph neural networks, we construct a 1-hop connectivity graph by connecting defenders within communication range $r_c=1$. Given that the default radius is $R=1$, we foresee that three neighboring agents within 1-hop would provide a wide sensing region for the defenders. Accordingly, we assume that communications occur in real-time with $N_D^f=3$. Each defender gathers information as input features that consist of $N_A^f=10$ closest intruder positions and $N_D^f=3$ closest defender positions. The used parameters are summarized in Table~\ref{tab:3}. 

%%%%%%%%%%%%%%%%%%%%     Table 1     %%%%%%%%%%%%%%%%%%%%%     
\begin{table}[t!]
\begin{center}
\begin{tabular}{c | c c}
\hline
Parameter name & Symbol & Value\\
\hline
Capturing distance & $\epsilon$ & 0.02\\
Field of view & $FOV$ & $\pi$ \\
Number of intruder features & $\mathit{N}_A^f$ & 10\\
Number of defender features & $\mathit{N}_D^f$ & 3\\
Communication range & $r_c$ & 1\\
Default team size & $N_{def}$ & 10 \\
\hline
\end{tabular}
\end{center}
\caption{Parameter setup in implementing graph neural networks, } 
\label{tab:3}
\end{table}
%%%%%%%%%%%%%%%%%%%%%%%%%%%%%%%%%%%%%%%%%%%%%%%%%%%%%%%%%% 

\subsubsection{Implementation Details}\label{sec:implementation}
The experiments are conducted using a 12-core 3.50GHz i9-9920X CPU and an Nvidia GeForce RTX 2080 Ti GPU. We implement the proposed networks using PyTorch v1.10.1~\citep{paszke2019pytorch} accelerated with Cuda v10.2 APIs. We use the Adam optimizer with a momentum of 0.5. The learning rate is scheduled to decay from $5\times10^{-3}$ to $10^{-6}$ within 1500 epochs with batch size 64, using cosine annealing. We choose these hyperparameters for the best performance.

%%%%%%%%%%%%%%%%%%%%%%%%%%%%%%%%%%%%%%%%%%%%%%%%%%%%%%%%%%%%%%%%%%%%%%%%%%%%%%%%
\section{Experiments}
\label{sec:experiments}

\subsection{Datasets}\label{subsec:dataset}
We evaluate our decentralized networks using imitation learning where the expert assignment policy is the maximum matching. The perimeter is a hemisphere with a radius $R$, which is defined by $R = \sqrt{N/N_{def}}$ where $N$ is team size and $N_{def}$ is a default team size. Since running the maximum matching is very expensive at large scales (e.g. $N>10$), we set the default team size $N_{def}=10$. In this way, $R$ also represents the scale of the game; for instance when $N=40$, $R$ becomes 2, which indicates that the scale of the problem's setting is doubled compared to the setting when $R=1$. Given the team size $N=10$, our experimental arena has a dimension of $10\times 10\times 1$ m. In offline, we randomly sample 10 million examples of defender's local perception $\mc Z_i$ and find corresponding $\mc G$ and $\mc L^g$ to prepare the dataset, which is divided into a training set (60\%), a validation set (20\%), and a testing set (20\%).

\subsection{Metrics}\label{subsec:metrics}
We are mainly interested in the percentage of intruders caught (i.e., number of captures/total number of intruders). At small scales (e.g. $N\leq 10$), an expert policy (i.e., the maximum matching) can be run and a direct comparison between the expert policy and our policy is available. At large scales (e.g. $N>10$), the maximum matching is too expensive to run. Thus we compare our algorithm with other baseline approaches: \textit{greedy}, \textit{random}, and \textit{mlp}, which will be explained in Section~\ref{subsec:compared}. To observe the scalability on small and large scales, we run a total of five different algorithms for each scale: \textit{expert}, \textit{gnn}, \textit{greedy}, \textit{random}, and \textit{mlp}. In all cases, we compute the \textit{absolute accuracy}, which is defined by the number of captures divided by the team size, to verify if our network can be generalized to any team size. Furthermore, we also calculate the \textit{comparative accuracy}, defined as the ratio of the number of captures by \textit{gnn} to the number of captures by another algorithm, to observe comparative results.

\subsection{Compared Algorithms}\label{subsec:compared}
In baseline algorithms, defenders do not communicate their “intentions” of which intruders would be captured by which neighboring defenders for a fair comparison since GNN does not share such information either. For the GNN framework, each defender perceives nearby intruders, and the relative positions of perceived intruders, not the “intentions,” are shared by GNN through communications. The power of the GNNs is to learn these “intentions” implicitly via K-hop communications. That way, the decentralized decision-making (i.e., for both GNN and baselines) may allow multiple defenders to aim to capture the same intruder while the centralized planner knows the “intentions” of all the defenders and would avoid such a scenario.

\subsubsection{Greedy} The greedy algorithm can be run in polynomial time and thus becomes a good candidate algorithm to be compared with our approach using GNN. For a fair comparison, we run a decentralized greedy algorithm based on local perception $\mc Z_i$ of $D_i$. We enable $K$-hop neighboring communications so that the sensible region of a defender is expanded as if the networking channels of GNN are active. The defender $D_i$ computes the payoff $p_{ij}$ (see Section~\ref{sec:payoff}) based on any sensible intruder $A_j$ and greedily chooses an assignment that minimizes the payoff $p_{ij}$. 

\subsubsection{Random} The random algorithm is similar to the greedy algorithm in that the $K$-hop neighboring communications are enabled for the expanded perception. Among sensible intruders, a defender $D_i$ randomly picks an intruder to determine the assignment.

\subsubsection{MLP} For the MLP algorithm, we only train the current MLP of our proposed framework in isolation by excluding the GNN module. By comparing our GNN framework to this algorithm, we can observe if the GNN gives any improvement.

\subsection{Results}\label{subsec:results}
We run the perimeter defense game in various scenarios with different team sizes and initial configurations to evaluate the performance of the learned networks. In particular, we conduct the experiments at small ($N\leq 10$) and large ($N>10$) scales. The snapshots of the simulated perimeter defense game in top view with our proposed networks for different team sizes are shown in Figure~\ref{fig:game}. The perimeter, defender state, intruder state, and breaching point are marked in green, blue, red, and yellow, respectively. We observe that intruders try to reach the perimeter. Given the defender-intruder matches, the intruders execute their respective optimal strategies to move towards the optimal breaching points (see Section~\ref{sec:nash}). If an intruder successfully reaches it without being captured by any defender, the intruder is consumed and leaves a marker called ``Intrusion". If an intruder fails and is intercepted by a defender, both agents are consumed and leave a marker called ``Capture". The points on the perimeter aimed by intruders are marked as ``Breaching point". In all runs, the game ends at \textit{terminal time} $T_f$ when all the intruders are consumed. See the supplemental video for more results. 

\begin{subfigure}[t!]
    \begin{minipage}[b]{0.32\linewidth}
        \centering   
        \includegraphics[height=4.1003293cm]{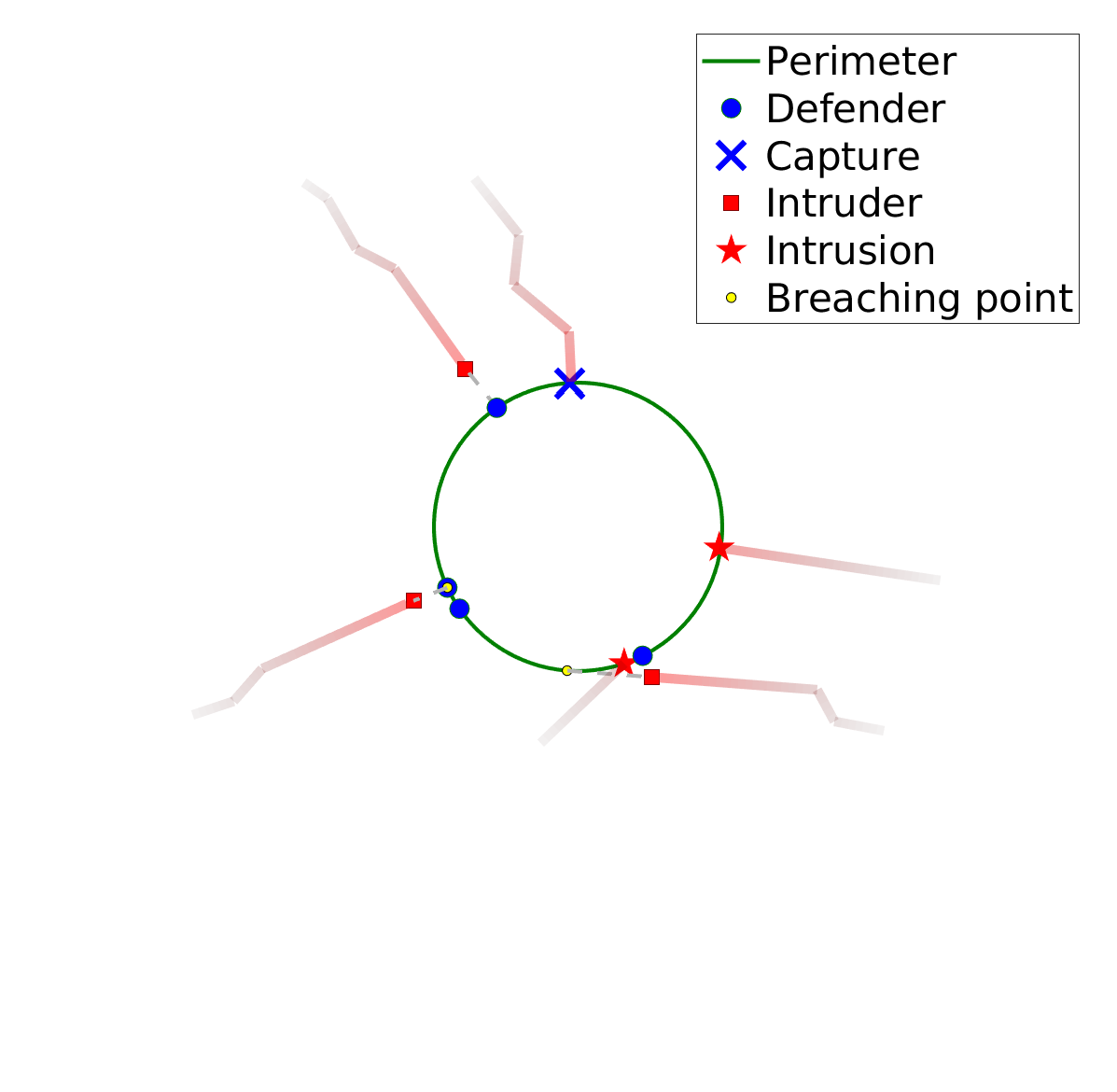}
        \caption{6 vs. 6}
    \end{minipage}
    \begin{minipage}[b]{0.32\linewidth}
        \centering
        \includegraphics[width=4.1003293cm]{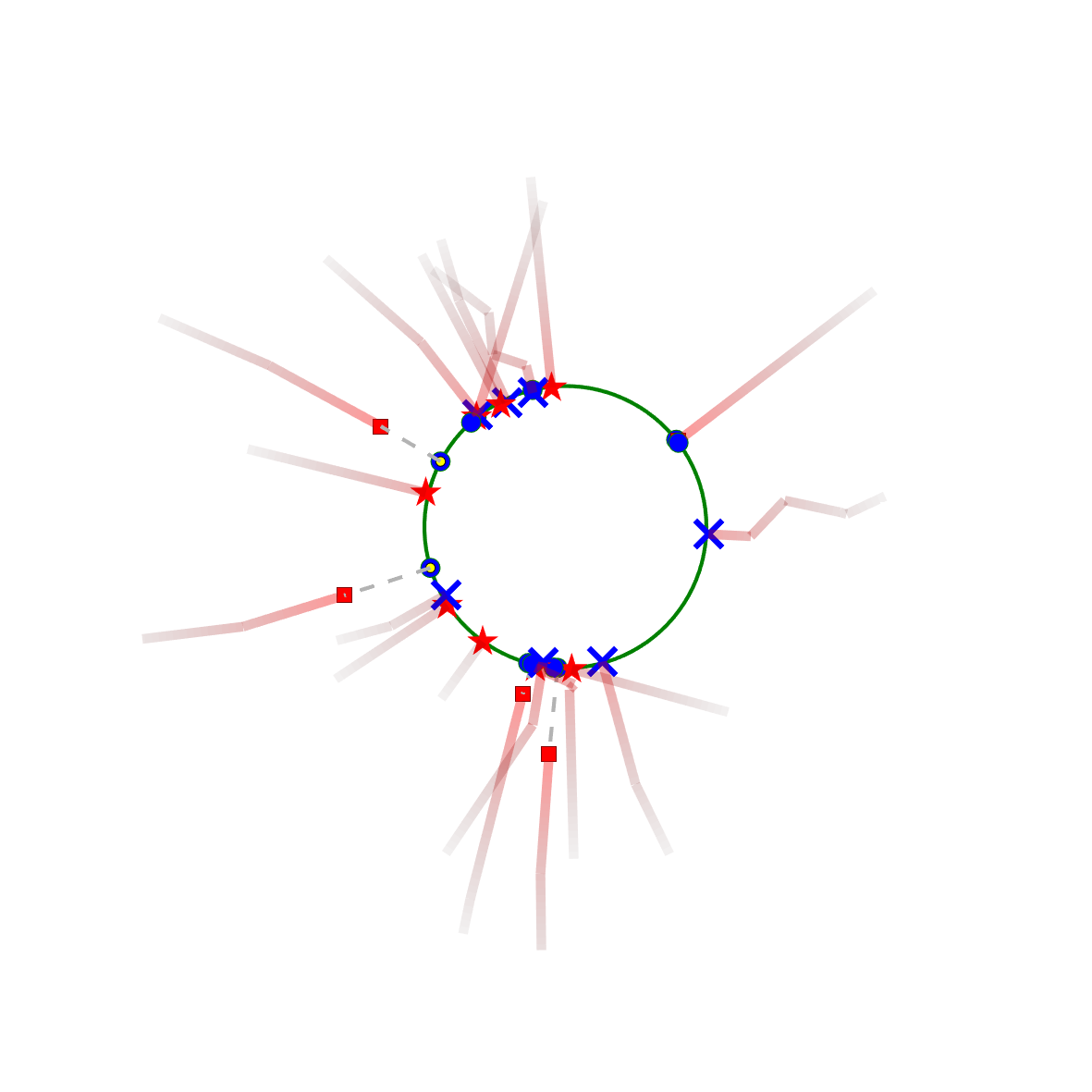}
        \caption{20 vs. 20}
    \end{minipage}
    \begin{minipage}[b]{0.32\linewidth}
        \centering
        \includegraphics[width=4.1003293cm]{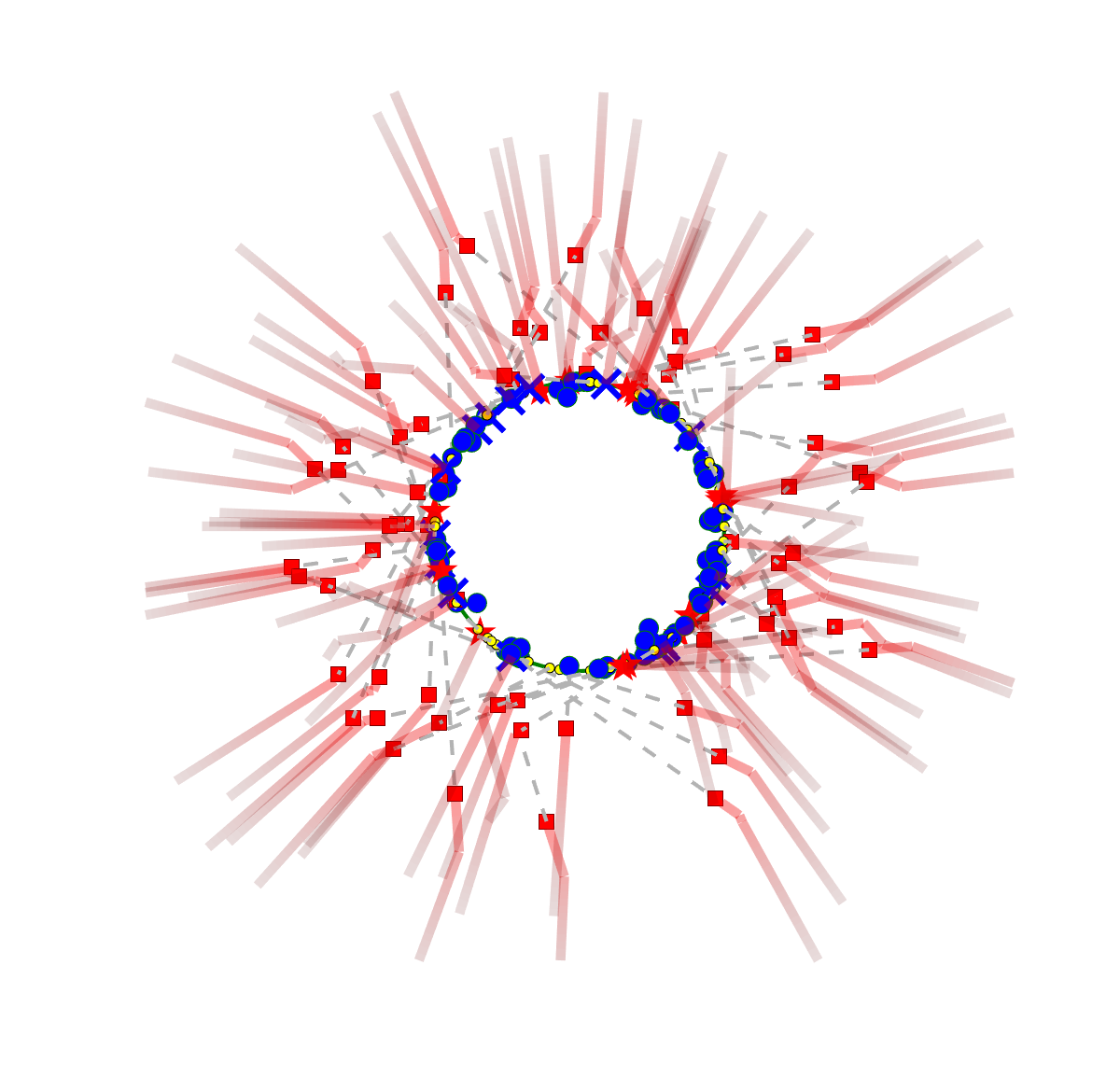}
        \caption{100 vs. 100}
    \end{minipage}
\setcounter{subfigure}{-1}
    \caption{\textbf{(A)-(C)} Snapshots of simulated perimeter defense in top view using the proposed method \textit{gnn} for three different team sizes.}
    \label{fig:game}
\end{subfigure}

As mentioned in Section~\ref{subsec:dataset}, we run the five algorithms \textit{expert}, \textit{gnn}, \textit{greedy}, \textit{random}, and \textit{mlp} at small scales, and run \textit{gnn}, \textit{greedy}, \textit{random}, and \textit{mlp} in large scales. As an instance, the snapshots of simulated 20 vs. 20 perimeter defense game in top view at terminal time $T_f$ using the four algorithms are displayed in Figure~\ref{fig:endgame}. The four subfigures (a)-(d) show that these algorithms exhibit different performance although the game begins with the same initial configuration in all cases. The number of captures by these algorithms \textit{gnn}, \textit{greedy}, \textit{random}, and \textit{mlp} are 12, 11, 10, 7, respectively.

\setcounter{figure}{6}
\setcounter{subfigure}{0}
\begin{subfigure}[b!]
    \begin{minipage}[b]{0.245\linewidth}
        \centering  
        \includegraphics[height=3.07cm]{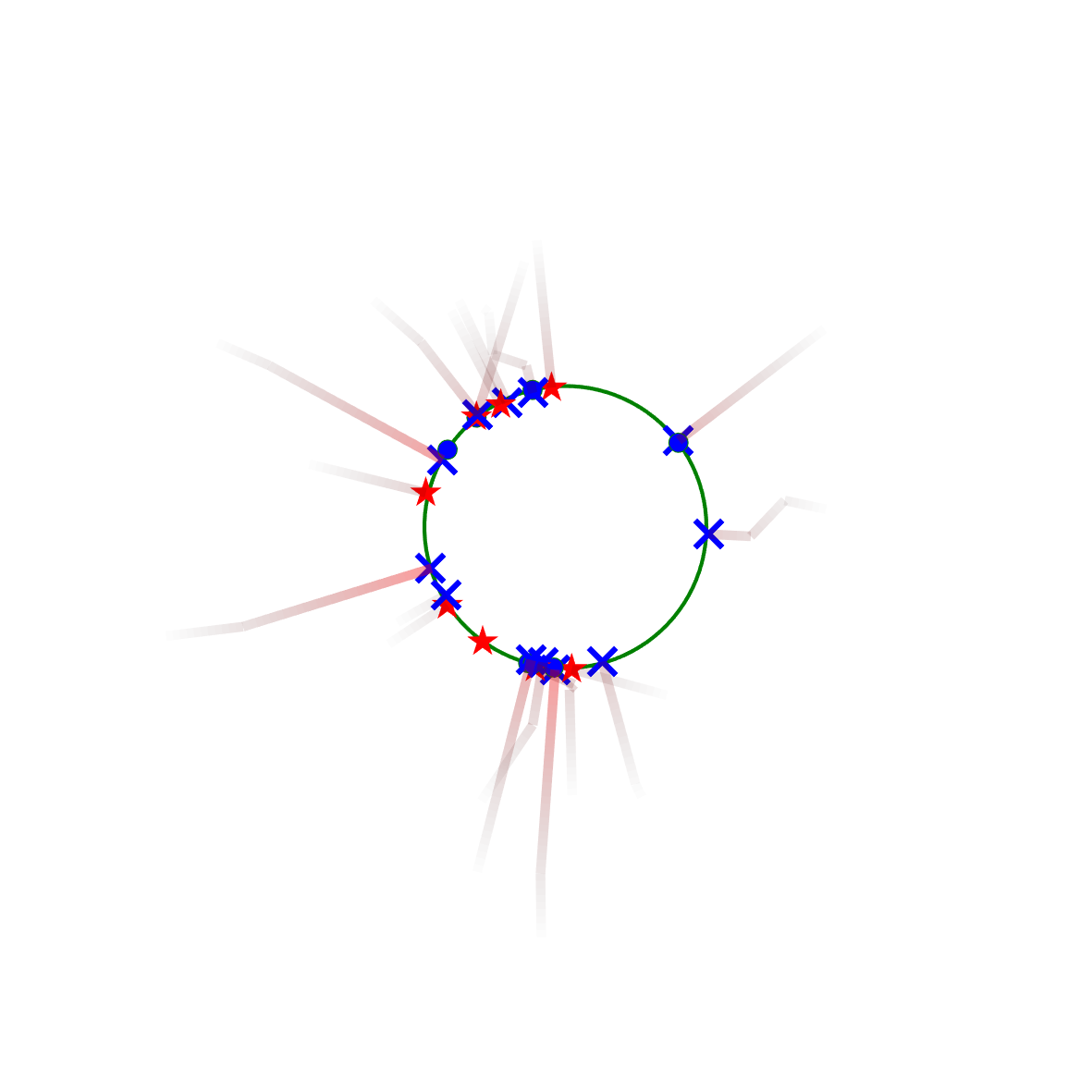}
        \caption{\textit{gnn}}
    \end{minipage}
    \begin{minipage}[b]{0.245\linewidth}
        \centering
        \includegraphics[height=3.07cm]{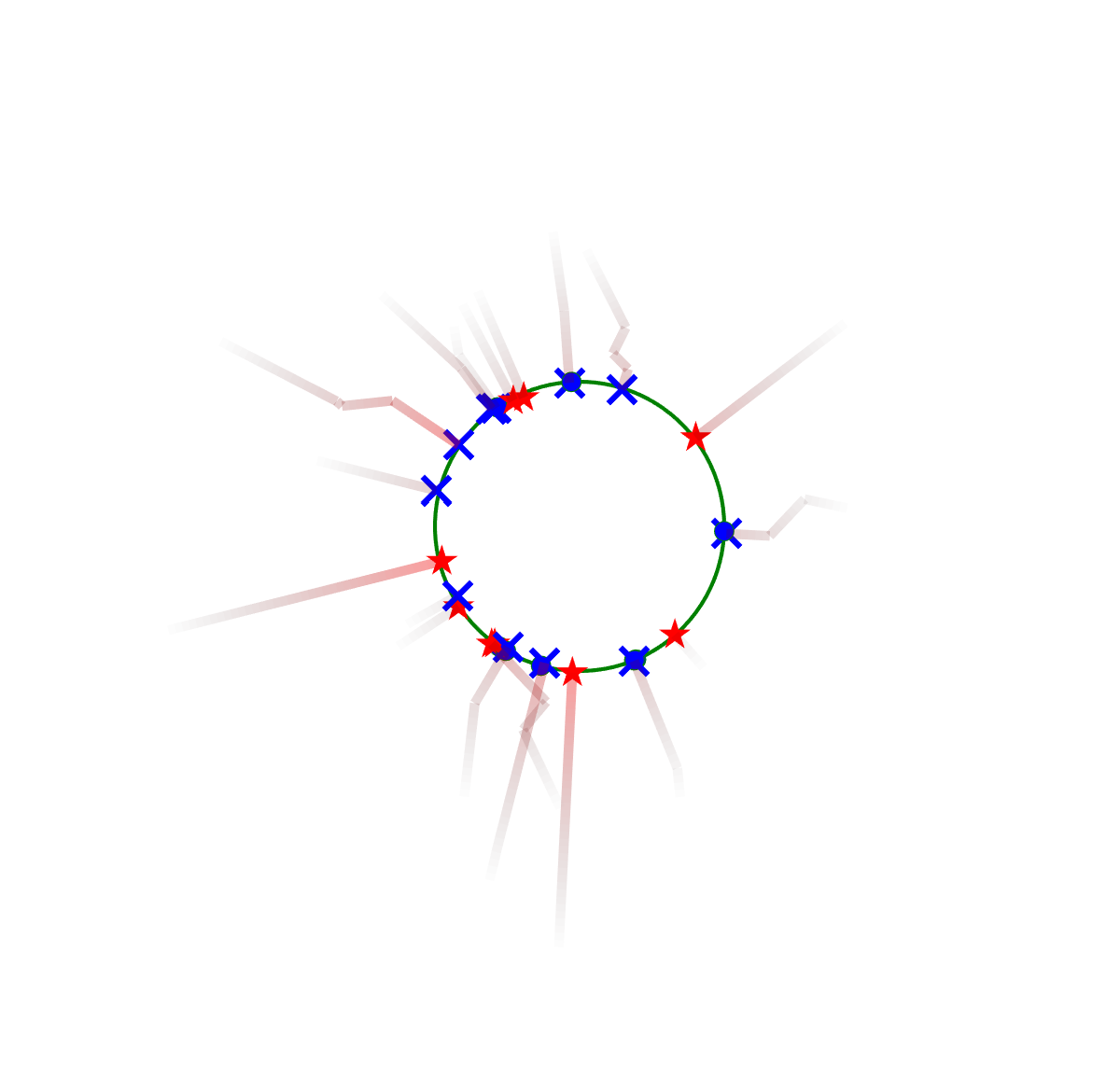}
        \caption{\textit{greedy}}
    \end{minipage}
    \begin{minipage}[b]{0.245\linewidth}
        \centering
        \includegraphics[height=3.07cm]{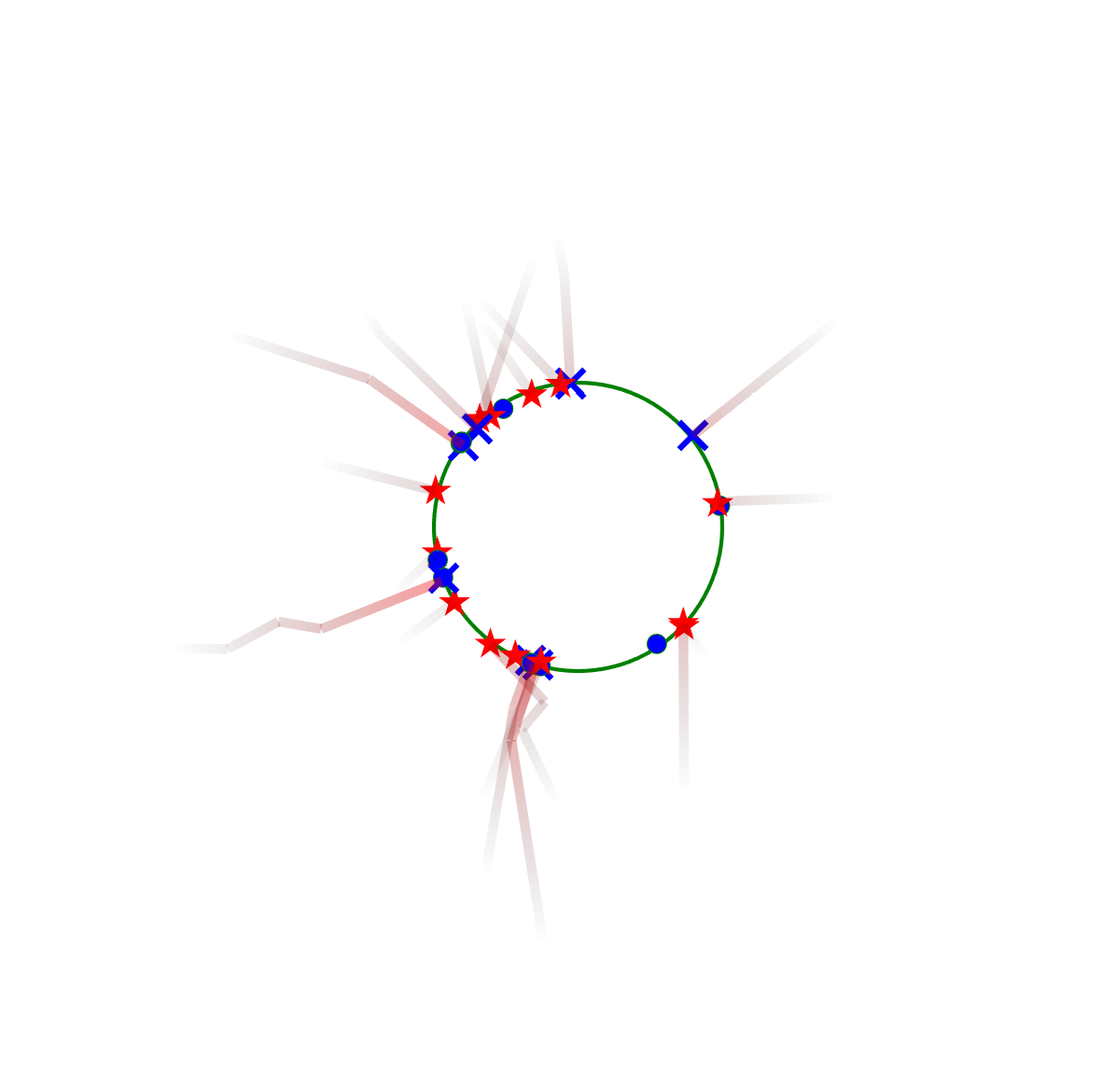}
        \caption{\textit{random}}
    \end{minipage}
    \begin{minipage}[b]{0.245\linewidth}
        \centering
        \includegraphics[height=3.07cm]{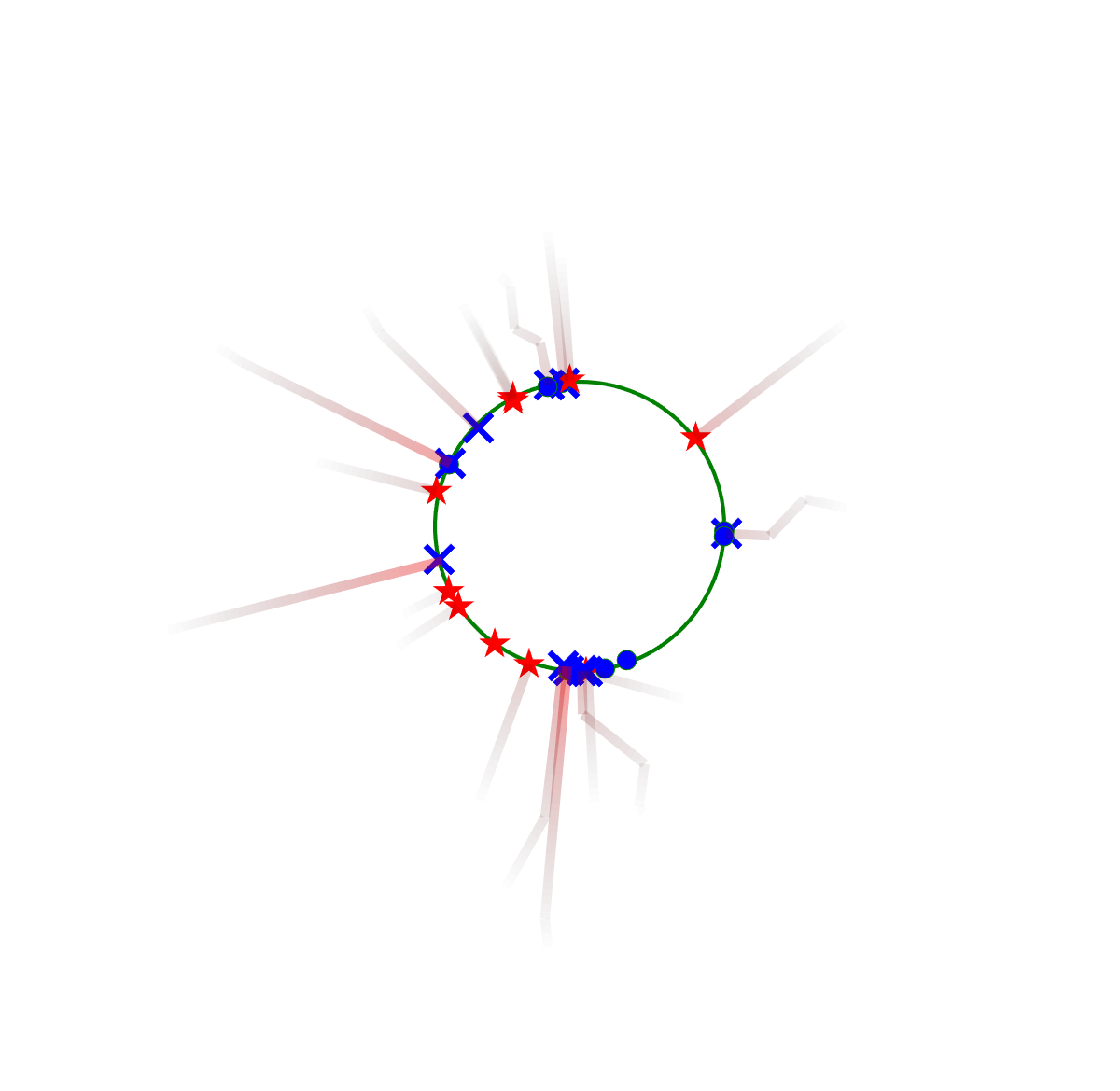}
        \caption{\textit{mlp}}
    \end{minipage}
\setcounter{subfigure}{-1}
    \caption{\textbf{(A)-(D)} Snapshots of simulated 20 vs. 20 perimeter defense game in top view at terminal time $T_f$ using the four algorithms \textit{gnn}, \textit{greedy}, \textit{random}, and \textit{mlp}. The number of captures using these algorithms are 12, 11, 10, and 7, respectively.}
    \label{fig:endgame}
\end{subfigure}

\begin{figure}[h!]
    \centering
    \includegraphics[width=0.495\textwidth]{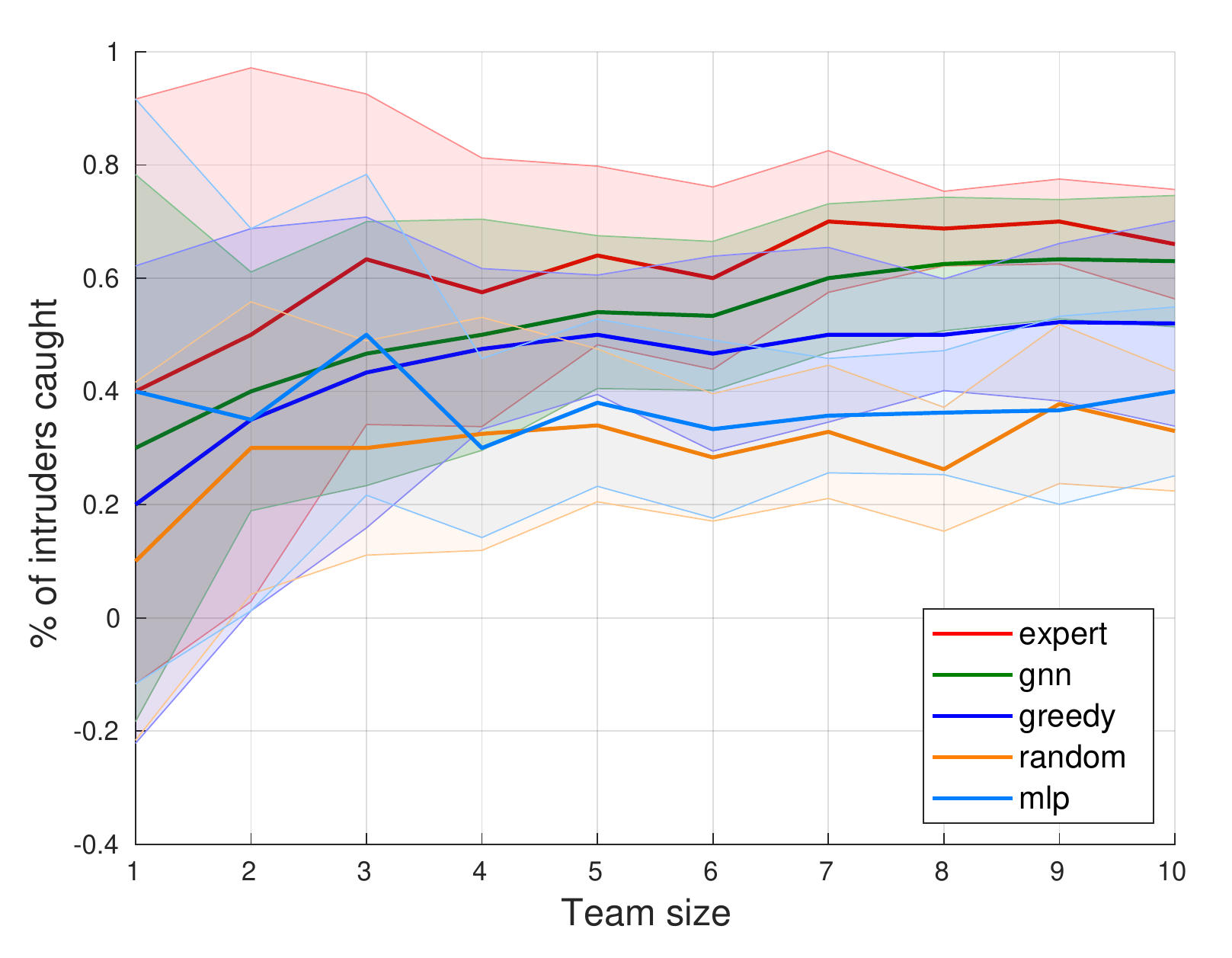}
    \includegraphics[width=0.49\textwidth]{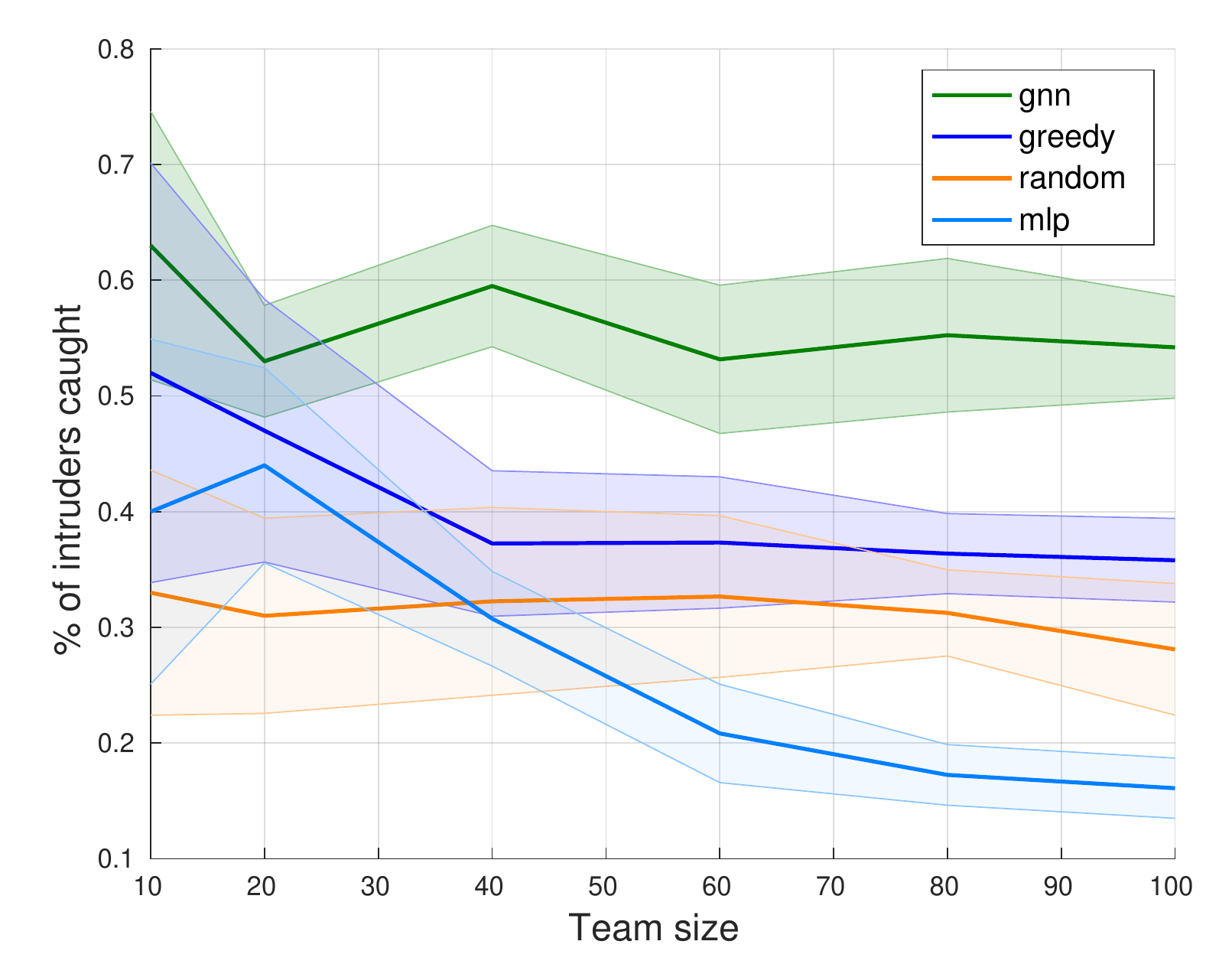}
    \caption{Percentage of intruders caught (average and standard deviation over 10 trials) by different algorithms on small ($N\leq10$) and large ($N>10$) scales.}
    \label{fig:scale}
\end{figure}

The overall results of the percentage of intruders caught by each of these methods are depicted in Figure~\ref{fig:scale}. It is observed that \textit{gnn} outperforms other baselines in all cases, and performs close to \textit{expert} at the small scales. In particular, given that our default team size $N_{def}$ is 10, the performance of our proposed algorithm stays competitive with that of the expert policy near $N=10$.

At large scales, the percentage of captures by \textit{gnn} stays constant, which indicates that the trained network can be well generalized to the large scales even if the training has been performed at the small scale. The percentage of captures by \textit{greedy} also seems constant but performs much worse than \textit{gnn} as the team size gets large. At small scales, only a few combinations are available in matching defender-intruder pairs and thus the \textit{greedy} algorithm would perform similarly to the expert algorithm. As the number of agents increases, the number of possible matching increases exponentially so the \textit{greedy} algorithm performs worse since the problem complexity gets much higher. The \textit{random} approach performs worse than all other algorithms at small scales, but the \textit{mlp} begins to perform worse than the \textit{random} when the team size increases over 40. This tendency tells that the policy trained only with MLP cannot be scalable at large scales. Since the training is done with 10 agents, it is optimal near $N=10$, but the \textit{mlp} cannot work at larger scales and even performs worse than the \textit{random} algorithm. It is confirmed that the GNN added to the MLP significantly improves the performance. Overall, compared to other algorithms, \textit{gnn} performs better at large scales than at small scales, which validates that GNN helps the network become scalable.

To quantitatively evaluate the proposed method, we report the \textit{absolute accuracy} and \textit{comparative accuracy} (defined in Section~\ref{subsec:metrics}) in Table~\ref{tab:1} and Table~\ref{tab:2}. As expected, the absolute accuracy reaches the maximum when team size approaches $N=10$. The overall values of the absolute accuracy are fairly consistent except when $N=2$. We conjecture that there may not be much information shared by the two defenders and there could be no sensible intruders at all based on initial configurations.

%%%%%%%%%%%%%%%%%%%%     Table 2     %%%%%%%%%%%%%%%%%%%%%     
\begin{table}[h!]
\begin{center}
\begin{tabular}{c | c c c c c}
\hline
Team Size & 2 & 4 & 6 & 8 & 10\\
\hline
Absolute accuracy (gnn vs. N) & 0.40 & 0.50 & 0.53 & 0.63 & 0.63\\
Comparative accuracy (gnn vs. expert) & 0.80 & 0.87 & 0.89 & 0.91 & 0.95\\
Comparative accuracy (gnn vs. greedy) & 1.14 & 1.05 & 1.14 & 1.25 & 1.21\\
Comparative accuracy (gnn vs. random) & 1.33 & 1.54 & 1.88 & 2.38 & 1.91\\
Comparative accuracy (gnn vs. mlp) & 1.14 & 1.67 & 1.60 & 1.72 & 1.58\\
\hline
\end{tabular}
\end{center}
\caption{Accuracy for small scales} 
\label{tab:1}
\end{table}
%%%%%%%%%%%%%%%%%%%%%%%%%%%%%%%%%%%%%%%%%%%%%%%%%%%%%%%%%%     

%%%%%%%%%%%%%%%%%%%%     Table 3     %%%%%%%%%%%%%%%%%%%%%     
\begin{table}[h!]
\begin{center}
\begin{tabular}{c | c c c c c}
\hline
Team Size & 20 & 40 & 60 & 80 & 100\\
\hline
Absolute accuracy (gnn vs. N) & 0.53 & 0.59 & 0.53 & 0.55 & 0.54\\
Comparative accuracy (gnn vs. greedy) & 1.13 & 1.59 & 1.42 & 1.52 & 1.51\\
Comparative accuracy (gnn vs. random) & 1.71 & 1.85 & 1.63 & 1.77 & 1.93\\
Comparative accuracy (gnn vs. mlp) & 1.20 & 1.94 & 2.55 & 3.20 & 3.37\\
\hline
\end{tabular}
\end{center}
\caption{Accuracy for large scales}
\label{tab:2}
\end{table}
%%%%%%%%%%%%%%%%%%%%%%%%%%%%%%%%%%%%%%%%%%%%%%%%%%%%%%%%%%  

The comparative accuracy between \textit{gnn} and \textit{expert} shows that our trained policy gets much closer to the expert policy as $N$ approaches 10, and we expect the performance of \textit{gnn} to be close to that of \textit{expert} even at the large scales. The comparative accuracy between \textit{gnn} and other baselines shows that our trained networks perform much better than baseline algorithms at the large scales ($N\geq 40$) with an average of 1.5 times more captures. The comparative accuracy between \textit{gnn} and \textit{random} is somewhat noisy throughout the team size due to the nature of randomness, but we observe that our policy can outperform random policy with an average of 1.8 times more captures at small and large scales. We observe that \textit{mlp} performs much worse than other algorithms at large scales.

Based on the comparisons, we demonstrate that our proposed networks, which are trained at a small scale, can generalize to large scales. Intuitively, one may think that \textit{greedy} would perform the best in a decentralized setting since each defender does its best to minimize the \textit{value of the game} (defined in Eq.~\ref{eqn:V}). However, we can infer that  \textit{greedy}  does not know the intentions of nearby defenders (e.g. which intruders to capture) so it cannot achieve the performance close to the centralized expert algorithm. Our method implements graph neural networks to exchange the information of nearby defenders, which perceive their local features, to plan the final actions of the defender team; therefore, implicit information of where the nearby defenders are likely to move is transmitted to each neighboring defender. Since the centralized expert policy knows all the intentions of defenders, our GNN-based policy learns the intention through communication channels. The collaboration among the defender team is the key for our \textit{gnn} to outperform \textit{greedy} approach. These results validate that the implemented GNNs are ideal for our problem with the properties of the decentralized communication that captures the neighboring interactions and transferability that allows for generalization to unseen scenarios.

\subsection{Further Analysis}

\subsubsection{Performance vs. Number of expert demonstrations}\label{sec:training}
\begin{figure}[t!]
    \centering
    \includegraphics[width=0.73\textwidth]{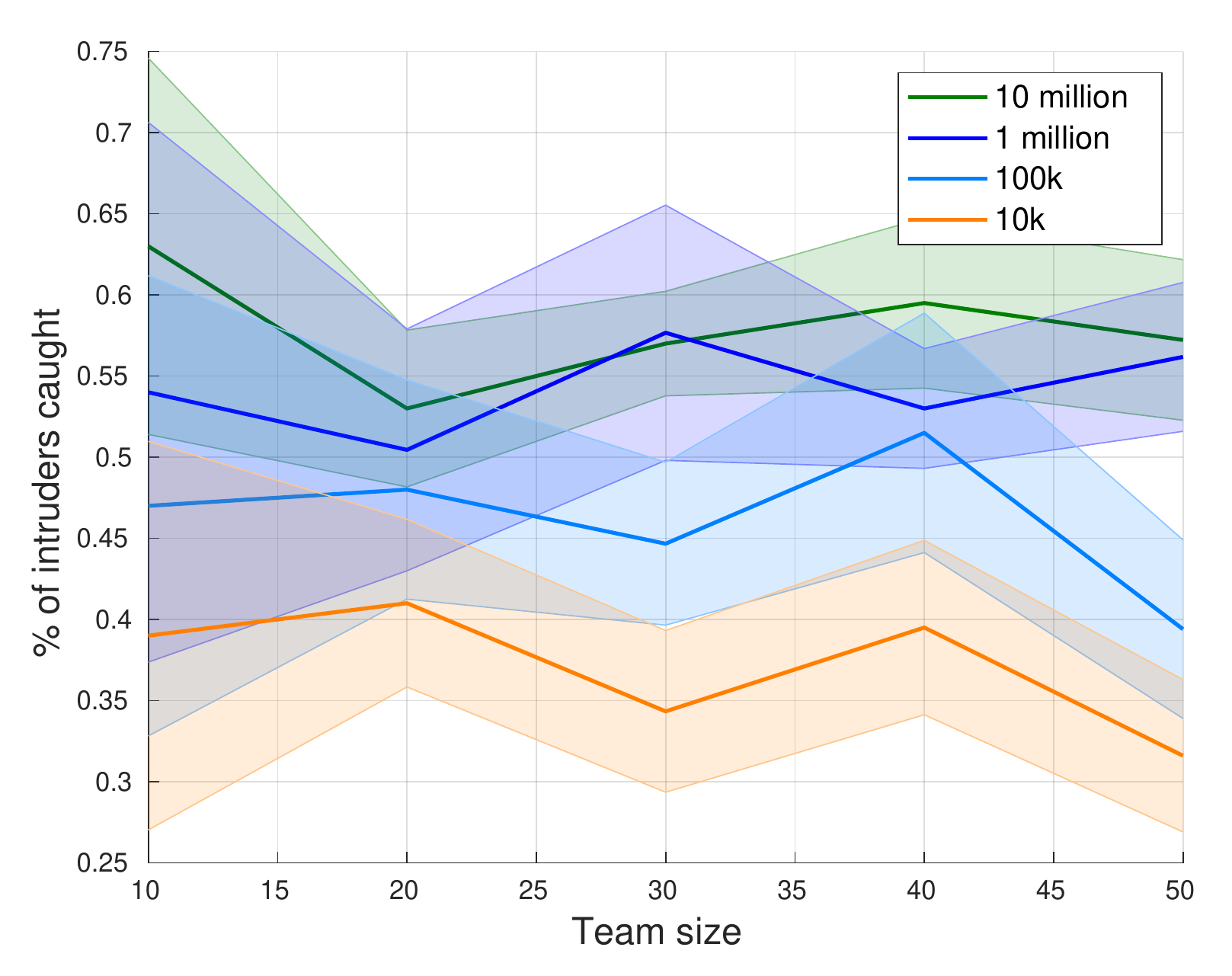}
    \caption{Sample efficiency with different numbers of expert demonstrations.}
    \label{fig:training}
\end{figure}

To analyze the algorithm performance, we have trained our GNN-based architecture with a different number of expert demonstrations (e.g., 10 million, 1 million, 100k, and 10k). The percentage of intruders caught (average and standard deviation over 10 trials) on team size $10 \leq N \leq 50$ are shown in Figure~\ref{fig:training}. The plot validates that our proposed network learns better with more demonstrations.

\subsubsection{Performance vs. Perimeter radius}
\begin{figure}[t!]
    \centering
    \includegraphics[width=0.7\textwidth]{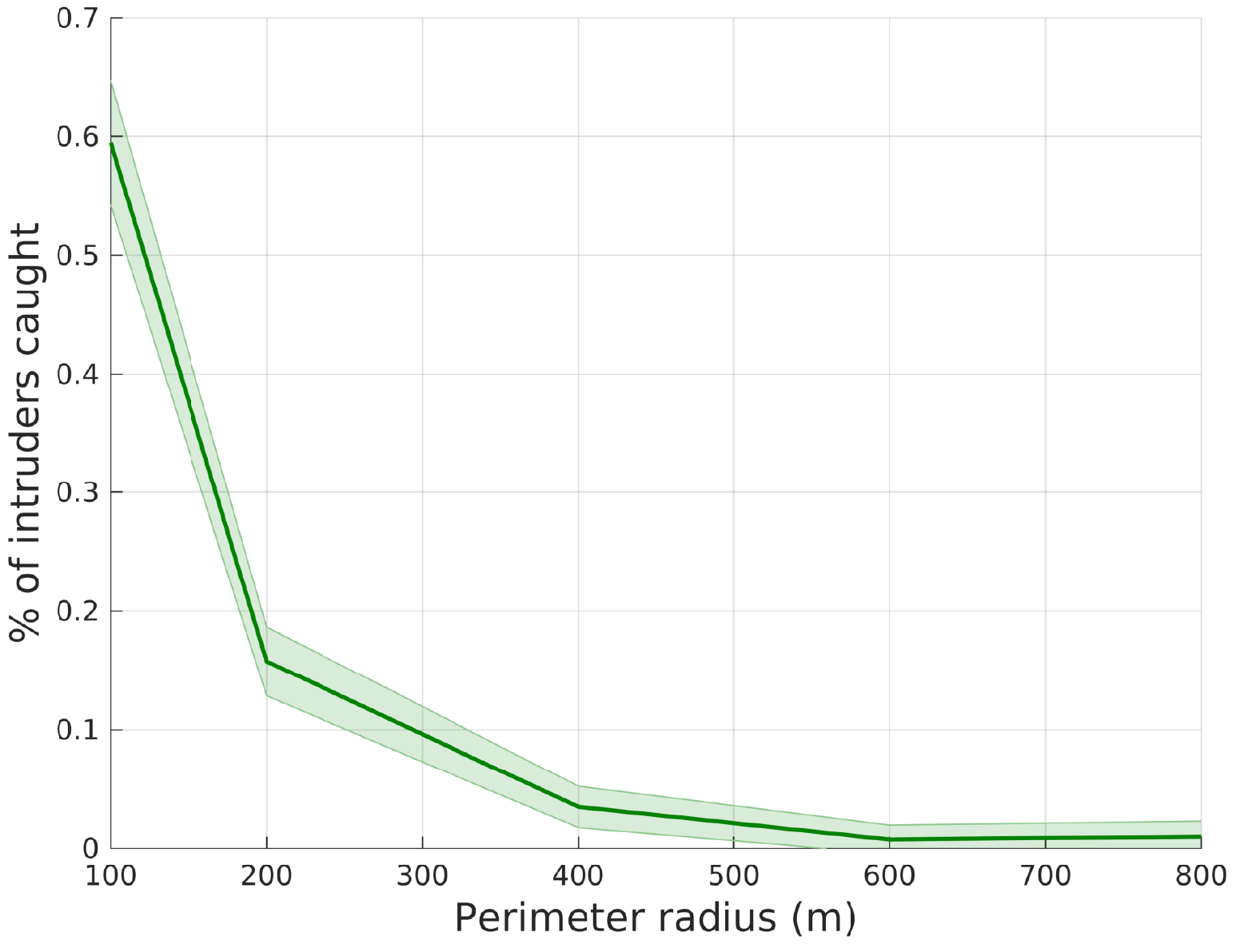}
    \caption{Percentage of intruders caught with various perimeter radii.}
    \label{fig:radius}
\end{figure}

We have tested the GNN-based proposed method with different perimeter radii. Intuitively, given the fixed number of agents, increasing the radius may lead to a failure in the defense system. We set the default team size of defenders as 40 and increase the perimeter radius until the percentage of intruders caught converges to zero. As shown in Figure~\ref{fig:radius}, the percentage decreases as the radius changes from 100m to 800m, converging to zero. 

\subsubsection{Performance vs. Number of intruders sensed}
\begin{figure}[h!]
    \centering
    \includegraphics[width=0.7\textwidth]{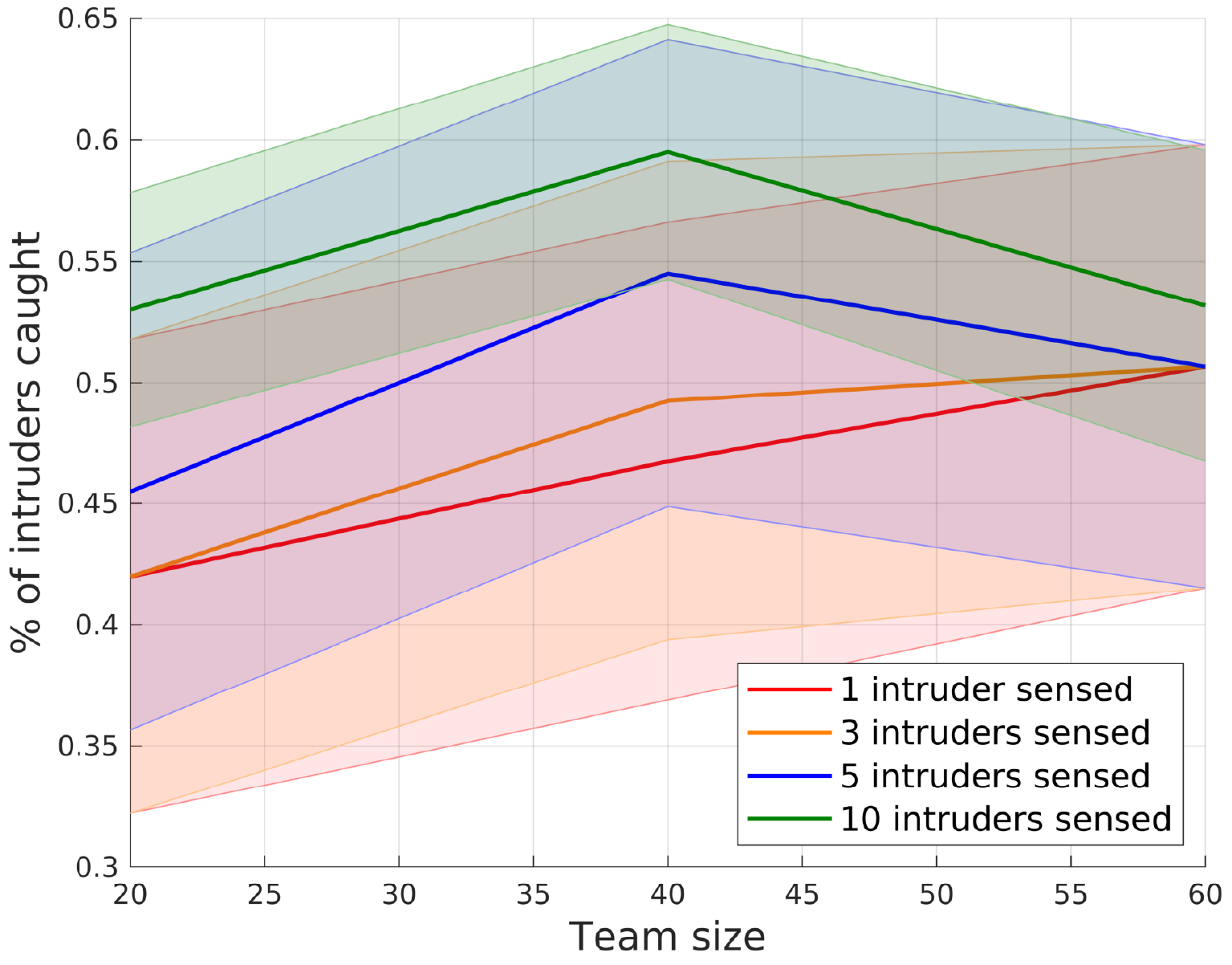}
    \caption{Percentage of intruders caught with different numbers of intruders sensed.}
    \label{fig:sensed}
\end{figure}

The performance of our GNN-based approach with different numbers of intruder (e.g., $N_A^f$) sensed is shown in Figure~\ref{fig:sensed}. We have run the experiments with $N_A^f$ as 1, 3, 5, and 10 since no ground truth expert policy is available to generate the training data for numbers larger than 10. We observe that the more intruder features are sensed, the better performances are shown. Further, the performance discrepancy tends to be smaller as the team size gets bigger. For some team size (e.g., 40), higher $N_A^f$ performs much better, but this is expected based on the initial configuration of the game. For instance, if the initial configuration is very sparse, a defender will benefit from higher $N_A^f$, and the percentage of intruders caught will be higher.

\subsection{Limitations}\label{subsec:limiations}
As perimeter defense is a relatively new field of research, this work has underlying limiting assumptions. In the problem formulation, we assume the robots are point particles. Accordingly, we assume optimal trajectories obey first-order assumptions. There is a preliminary work~\citep{lee2021defending} to bridge the gap between the point particle assumptions and three-dimensional robots for one-on-one hemisphere perimeter defense, and we hope to extend the idea of this work to our multi-agent perimeter defense problem in the future. Another limitation is that there is no available expert policy, which can be compared with our proposed method, at large scales. Running the maximum matching algorithm is very expensive at large scales, so we compare our GNN-based algorithm with other baseline methods. Although the consistent performances of tested algorithms along different scales confirm that our trained networks can be generalized to large scales, we hope to explore another algorithm that can be used as an expert policy at large scales to replace the maximum matching. One consideration is utilizing reinforcement learning since the algorithm performance at large scales will be available.

%%%%%%%%%%%%%%%%%%%%%%%%%%%%%%%%%%%%%%%%%%%%%%%%%%%%%%%%%%%%%%%%%%%%%%%%%%%%%%%%
\section{Conclusion}\label{sec:conclusion}
This paper proposes a novel framework that employs graph neural networks to solve the decentralized multi-agent perimeter defense problem. Our learning framework takes the defenders' local perceptions and the communication graph as inputs and returns actions to maximize the number of captures for the defender team. We train deep networks supervised by an expert policy based on the maximum matching algorithm. To validate the proposed method, we run the perimeter defense game in different team sizes using five different algorithms: \textit{expert}, \textit{gnn}, \textit{greedy}, \textit{random}, and \textit{mlp}. We demonstrate that our GNN-based policy stays closer to the expert policy at small scales and the trained networks can generalize to large scales. 

One future work is to implement vision-based local sensing for the perception module, which would relax the assumptions of perfect state estimation. Realizing multi-agent perimeter defense with vision-based perception and communication within the defenders will be an end goal. Another future research direction is to find a centralized expert policy in multi-robot systems by utilizing reinforcement learning.

%%%%%%%%%%%%%%%%%%%%%%%%%%%%%%%%%%%%%%%%%%%%%%%%%%%%%%%%%%%%%%%%%%%%%%%%%%%%%%%%

\section*{Data Availability Statement}
The raw data supporting the conclusions of this article will be made available by the authors, without undue reservation.

\section*{Author Contributions}
EL, LZ, and VK contributed to conception and design of the study. EL and AR performed the statistical analysis. EL wrote the first draft of the manuscript. All authors contributed to manuscript revision, read, and approved the submitted version.

\section*{Acknowledgments}
We gratefully acknowledge the support from 
ARL DCIST CRA under Grant W911NF-17-2-0181,
NSF under Grants CCR-2112665, CNS-1446592, and EEC-1941529,
ONR under Grants N00014-20-1-2822 and N00014-20-S-B001, 
Qualcomm Research, 
NVIDIA,
Lockheed Martin,
and C-BRIC, a Semiconductor Research Corporation Joint University Microelectronics Program cosponsored by DARPA.

\section*{Conflict of Interest}
The authors declare that the research was conducted in the absence of any commercial or ﬁnancial relationships that could be construed as a potential conﬂict of interest.

% Perimeter defense is a relatively new field of research that has been explored recently. One particular challenge is that the high-dimensional perimeters add spatial and algorithmic complexities for defenders to execute their optimal strategies. Although many previous works considered engagements on a planar game space and derived optimal strategies in 2D motions, the extension towards high-dimensional areas is unavoidable for practical applications of perimeter defense games in real-world scenarios. This work aims to simplify high-dimensional problems by developing a framework where a team of defenders collaborates to defend the perimeter using decentralized strategies based on local perceptions and communications. Specifically, we explore learning-based approaches to learn policies by imitating expert algorithms such as the maximum matching algorithm, which is computationally intensive at large scales since this approach is combinatorial in nature and assumes global information. We utilize GNN as the learning paradigm and demonstrate that the trained network can perform close to the expert algorithm. GNNs have a decentralized communication architecture that captures the neighboring interactions and transferability, allowing for generalization to previously unseen scenarios. We demonstrate that our proposed GNN-based network can be generalized to large scales in solving multi-robot perimeter defense games.

\bibliographystyle{Frontiers-Harvard} %  Many Frontiers journals use the Harvard referencing system (Author-date), to find the style and resources for the journal you are submitting to: https://zendesk.frontiersin.org/hc/en-us/articles/360017860337-Frontiers-Reference-Styles-by-Journal. For Humanities and Social Sciences articles please include page numbers in the in-text citations 
\bibliography{frontiers}

\end{document}